\documentstyle[epsf]{mn}

\newcommand{\etal}{{ et al.\ }}
\newcommand{\mnras}{MNRAS}
\newcommand{\apj}{ApJ}
\newcommand{\apjs}{ApJS}

\title{Gas cooling in simulations of the formation of the
galaxy population}
\author[N. Yoshida, F. Stoehr, V. Springel \& S. D. M. White]
{Naoki Yoshida$^{1,2}$, Felix Stoehr$^1$, Volker Springel$^1$ and Simon D. M. White$^1$\\
$^1$ Max-Planck-Institut f\"{u}r Astrophysik, Karl-Schwarzschild-Str.1,
Garching bei M\"unchen, D85748 Germany\\
$^2$ Harvard-Smithsonian Center for Astrophysics, 60 Garden Street,
Cambridge MA 02138, USA}
\date{To appear in MNRAS, 2002}

\begin{document}
\maketitle

\begin{abstract}
We compare two techniques for following the cooling of gas and its 
condensation into galaxies within high resolution simulations of 
cosmologically representative regions. Both techniques 
treat the dark matter using $N$-body methods. One follows the gas using
smoothed
particle hydrodynamics (SPH) while the other uses simplified recipes
from semi-analytic (SA) models. We compare the
masses and locations predicted for dense knots of cold gas (the
`galaxies') when the two techniques are applied to evolution from
the same initial conditions and when the additional complications of
star formation and feedback are ignored. We find that above the
effective
resolution limit of the two techniques, they give very similar results
both for global quantities such as the total amount of cooled gas and
for the properties of individual `galaxies'. The SA technique has 
systematic uncertainties arising from the simplified cooling
model adopted, while details of the SPH implementation can produce
substantial systematic variations in the galaxy masses it predicts.
Nevertheless, for the best current SPH methods and the standard
assumptions of the SA model, systematic differences between the two 
techniques are remarkably small. The SA technique 
gives adequate predictions for the condensation of gas into 
`galaxies' at less than one percent of the computational cost of
obtaining similar results at comparable resolution using SPH.
\end{abstract}


\section{Introduction}
Gas-dynamical and radiative processes couple with gravitational
instability to play a crucial role in the formation of galaxies. 
In hierarchically clustering universes galaxies are formed via 
the dissipative cooling and condensation of gas within 
dark matter halos. The analytic understanding
of galaxy formation within this framework is well developed (e.g. White 
1996 and references therein). It is most practically embodied 
in the so-called `semi-analytic' (SA) models. These implement simple 
analytic treatments of baryonic processes within a model for the 
nonlinear development of the dark matter distribution to give Monte 
Carlo realisations of the evolution of the galaxy population (White 
\& Frenk 1991; Lacey \& Silk 1991; Kauffmann, 
White \& Guiderdoni 1993; Cole\etal 1994; Somerville \& Primack 1998).

A full description of galaxy formation requires treatment of many
complex physical processes. The huge dynamic range between the
scales on which individual stars form and those of a `typical'
region of the Universe makes it impossible to simulate all relevant
processes together. Some kind of `sub-grid' model must therefore be
adopted to describe the unresolved physics of star-formation and
feedback within a dynamical treatment of larger scale evolution.
Semi-analytic modelling adopts this philosophy not only on small
scales, but also on scales where the evolution can, at some 
computational cost, be simulated directly. In SA models the merging 
history of dark matter halos is taken as the principal  
dynamical input. The following additional key physical processes are 
then included:
(1) shock heating and virialisation of gas within halos,
(2) radiative cooling and condensation of this gas,
(3) star formation in the cold dense gas, 
(4) energy feedback from supernovae and stellar winds,
(5) metal enrichment, 
(6) stellar evolution, and
(7) galaxy merging. 
Early work by White \& Rees (1978) and White \& Frenk (1991) 
showed that regulation of star formation by feedback
is necessary to avoid substantial over-production of stars at early
times and in small objects, i.e. at least the first four of the above 
processes must be taken into account to obtain a realistic model.
More recent semi-analytic models have incorporated all these processes
in a consistent manner into Monte Carlo halo merger trees constructed
using the (extended) Press-Schechter 
theory (Kauffmann\etal 1993; Cole\etal 1994, 2000; Kauffmann,
Nusser \& Steinmetz 1997; Somerville \& Primack 1998; Benson\etal
1999). The output from such models has been tested against a wide
range of observational data, showing that many properties of the 
observed galaxy population can be successfully reproduced. A
recent development has been to replace the Monte Carlo merger
trees with trees constructed directly from the output of high
resolution $N$-body simulations (Kauffmann\etal 1999, hereafter
KCDW). This removes uncertainties related to the Press-Schechter halo
merger model and allows the evolution of the galaxy distribution to be
followed explicitly and in detail. With the high resolution of
Springel\etal (2001b, hereafter SWTK) it is also possible
to follow the merging of
the brighter galaxies explicitly, thus removing item (7) from the
list of processes to be treated by semi-analytic recipes. 

Although the models underlying these recipes 
are physically plausible and the resulting galaxy populations
are in reasonable agreement with observation,
the validity and the limitations of each model remain
unclear. Among the seven ingredients listed above, the first two
are well understood and can be incorporated in
hydrodynamic simulations in a direct and transparent manner.
Recently, Benson\etal (2001) studied the evolution of cooling gas
associated with galaxy formation both in smoothed particle
hydrodynamics (SPH) simulations and in a semi-analytic model
using Monte Carlo merger trees. Comparing the output of the two 
models, they found reasonable agreement in properties 
such as the global fractions of gas in cold, hot
and uncollapsed phases, and the amount of cold gas 
predicted in a dark halo as a function of its mass. Their results
suggest that the assumptions of semi-analytic models give an
appropriate description of the cooling of gas onto `protogalaxies'.
Since the Monte Carlo merger histories they assign
to each halo in their simulation are unrelated to its actual 
formation history, it is unclear, however, whether 
gas behaves as predicted by the SA model on an object-by-object 
basis. 

In this paper we address this issue
using a set of numerical simulations of rich galaxy cluster formation. 
We carry out two SPH simulations including radiative cooling, but excluding
star-formation and the related feedback. The simulations differ only
in the specific implementation of SPH employed. We compare the
`galaxy' populations of these simulations with those produced by
implementing a stripped-down SA model in which 
star-formation and feedback are also suppressed. This comparison
tests for differences arising from the very different treatments
of hydrodynamics and cooling, and is unaffected by differences
in the recipes which the SA and SPH approaches typically use for
star formation and feedback. Because both techniques follow the
dark matter distribution using the {\it same} $N$-body particle
representation, we are able to compare the galaxy
populations on an object-by-object basis.

The remainder of the paper is organised as follows. Sections 2 and 
3 describe our SPH simulations and present their results.
In Section 4, we explain our semi-analytic modelling
highlighting features included specifically
for the present study. We study and compare the effective resolution
of our SA and SPH techniques in Section 5.
In Section 6, we describe how we match the galaxies formed by
the two techniques, and we show how their masses compare.
We discuss the implications of these results in our concluding
Section 7.

\section{The $N$-body/SPH simulations}
We work with a flat $\Lambda$-dominated Cold Dark Matter universe,
with matter density $\Omega_{\rm m}=0.3$,
cosmological constant $\Omega_{\Lambda}=0.7$ and expansion rate
$H_{0}=70$km~s$^{-1}$Mpc$^{-1}$. We carry out further simulations
of the cluster studied by SWTK and Yoshida\etal
(2000a,b) which is the second most massive cluster in the GIF simulation
of KCDW. As in these earlier papers, we resimulate 
the cluster with a multi-mass technique in which only the region
immediately surrounding the object of interest is represented at
high resolution; the rest of the simulated volume is followed at
much lower resolution and serves merely to provide the correct
tidal gravitational field on the high resolution region.
Within this region we split each particle into
a dark matter particle of mass
$1.2\times 10^{10} h^{-1}M_{\odot}$ and a gas particle of mass
$2.1\times 10^{9} h^{-1}M_{\odot}$, and
we set the gravitational softening length to be 
$30h^{-1}$kpc. The simulations are carried out with the parallel tree 
$N$-body/SPH code GADGET (Springel, Yoshida \& White 2001a);
further details are presented in Yoshida (2001). 
During the simulations, we dump 50 snapshots of the particle data
spaced logarithmically in the cosmic expansion parameter from redshift 
$z=20$ to $z=0$. We use these outputs
for identifying and tracing galaxies, and also for constructing
dark matter halo merger trees for use in the semi-analytic simulation.
For the latter we throw out the gas particles and increase the mass
of each dark matter particle to $1.4\times 10^{10} h^{-1}M_{\odot}$.

We implement gas cooling in the SPH part of the code
in a similar manner to that in Katz, Weinberg \& Hernquist (1996,
hereafter KWH). As in KWH we compute the abundances of ionic species by
assuming collisional equilibrium for a gas of primordial composition
consisting of 76\% hydrogen and 24\% helium.  We call these simulations
`S0-SPH'. They include a time-dependent uniform UV 
background radiation, as in Dav\'{e} et al. (1999), 
but its effect, especially on the destruction of small objects, is 
very small at the resolution we are using (Weinberg, Katz, \& 
Hernquist 1997). Hence the inclusion of the UV heating is not
important for the analysis presented in this paper. We fix a minimum 
SPH smoothing length which prevents the gas smoothing
length from dropping below a quarter of the gravitational softening
length. We set the number of neighbouring particles in the SPH
smoothing kernel, $N_{\rm NGB}$, to be 32. This smoothing in the 
SPH simulation poses an effective resolution limit for gas cooling; 
gas cooling can take place efficiently only within objects that 
become dense enough and which contain enough gas particles.
We will discuss this resolution effect in more detail in Sections 4
and 5.

An important feature of our simulations is that we employ two
different implementations of SPH. One is an energy- 
and momentum-conserving scheme based on taking geometric means
of the pairwise hydrodynamic forces between neighboring particles;
it is very similar to that employed by KWH, Weinberg\etal (1997),
Dav\'{e}\etal (1999) and White, Hernquist \& Springel (2001)
among others. The other is a novel 
formulation of SPH recently suggested by Springel \& Hernquist (2001), 
which manifestly conserves momentum,  
energy and entropy in non-shock regions, even when adaptive smoothing 
lengths are employed. In a series of numerical tests, Springel \&
Hernquist showed that several
of the commonly employed formulations of SPH can give 
rise to an over-cooling phenomenon which results from spurious losses of 
specific entropy in poorly resolved cooling flows. This effect is 
particularly strong when the geometric symmetrisation of the 
hydrodynamical forces is employed. Our two simulations
highlight the extent to which such differences in the formulation
of SPH can affect the final result. In the following, we concentrate
on results from the simulation carried out with the new `entropy' 
scheme, but we stress the major points at which they differ from
those found using the geometric averaging technique. Note that while
we will often refer to the latter technique as the `conventional' 
SPH technique, it suffers from the overcooling problem more
severely than other `conventional' techniques which have also been
used in studies of galaxy formation (e.g. Benson\etal 2001). 
Pearce et al. (2001) found that ignoring cold dense particles
in the SPH density estimate for hot particles successfully
reduces the over-cooling of gas. 
Although we do not employ this phase-decoupling technique
in our SPH simulations, the problem
of density overestimates for hot particles in the vicinity of the 
condensed cold phase is much alleviated in the new entropy scheme 
(see Figure 8 in Springel \& Hernquist, 2002), so the effect of
the decoupling is expected to be small in our simulations.

\begin{figure}
\centering
\epsfxsize=\hsize\epsffile{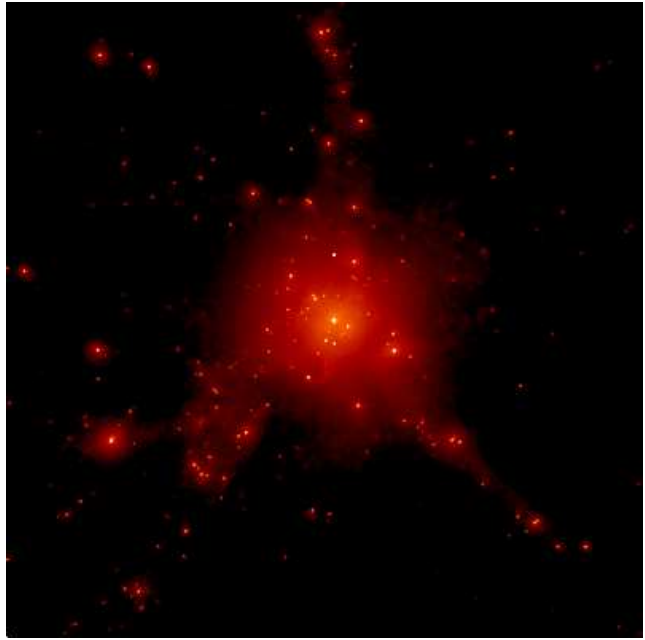}
\caption[The projected gas mass distribution.]
{The projected gas density distribution in the simulation
at $z=0$. The region shown is a cube of $15h^{-1}$Mpc on a side.}
\label{fig_cool1}
\end{figure}

We begin by presenting the basic properties of our simulated cluster.
The virial radius of the main cluster at the present epoch is $1.48
h^{-1}$Mpc, and 
the corresponding virial mass is  $7.7\times 10^{14} h^{-1}M_{\odot}$.
We define the virial radius $R_{\rm vir}$ as the radius of the sphere
centred on the most bound particle of the FOF group having 
overdensity 200 with respect to the critical density. The virial
mass $M_{\rm vir}$ is then the enclosed mass within $R_{\rm vir}$.
Figure~\ref{fig_cool1} shows the projected gas mass distribution
around the main cluster. Bright clumps seen in Figure~\ref{fig_cool1} 
are groups of cold gas particles, which we will later define as `galaxies'.
The diffuse component in the figure consists mainly of a shock-heated 
hot gas in and around the cluster. 

\begin{figure}
\centering
\epsfxsize=\hsize\epsffile{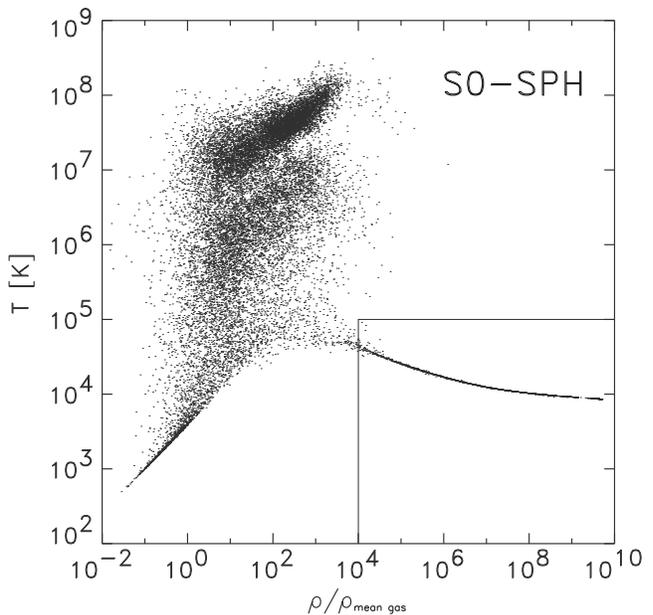}
\caption[Distribution of gas in the density-temperature plane at z=0.]
{The distribution of gas in the density-temperature plane at $z=0$.
Densities are normalised to the mean gas density.}
\label{fig_cool4}
\end{figure}

 Figure~\ref{fig_cool4} shows the distribution of gas in the
thermodynamic phase plane. The cold, dense gas particles 
appear clearly as a narrow tail in the plot, whereas the shock-heated
gas and the adiabatically cooled, unshocked gas appear in the upper and 
in the lower left portions, respectively.
The rectangular box approximately separates out
the gas particles in the `galaxy' phase. In practice 
we identify galaxies by linking the cold gas particles
using the standard FOF technique with a small linking length.
We describe our galaxy identification in the next section.

\section{Cold and hot gas in dark matter halos}
In this section we show the distribution and the properties 
of gas in the dark halos in the simulations.
We locate the dark halos 
by running a FOF group finder with linking parameter $b=0.164$.
We assigned virial radii and virial masses
to the dark halos in the manner described above.
We found 833 dark halos with virial mass larger than
$10^{11} h^{-1}M_{\odot}$ at the present epoch.  

\subsection{Gas mass fraction}
In Figure~\ref{fig_cool3} we plot the total gas mass and cold gas
fraction
against the host halo mass in the `entropy' version of the
S0-SPH simulation. Gas quantities are calculated within a sphere
of radius $R_{\rm vir}$. We define the cold gas fraction 
as the total mass of all the cold ($T < 100,000$ K) gas 
divided by the total gas mass within the halo. 
Three sets of plots for the outputs at z=0, 1, and 2,
are shown in Figure~\ref{fig_cool3}. 
In the top panels the solid line indicates
the global mean baryon fraction in the simulation, $f_{\rm b}$=0.15. 
A tight correlation is seen, showing that
the baryon fractions of the halos are very close to the universal
baryon fraction. At mass scales larger than $10^{12} h^{-1}M_{\odot}$ 
(corresponding about 80 dark matter particles), the scatter is very
small. At early times, there is a clear tendency for the gas fraction
to lie slightly above the global fraction. This is a consequence
of the rapid loss of pressure support behind the accretion shock when
cooling is efficient. 

The bottom panels in Figure~\ref{fig_cool3} show that the cold
gas fraction is a decreasing function of halo mass and shows
relatively little scatter above halo masses of $10^{12}
h^{-1}M_{\odot}$. In this regime the
cold gas mass fraction at given halo mass becomes progressively 
{\it lower} at lower redshift. This behaviour is expected as a 
result of the decrease in cooling efficiency at lower redshift
produced by the lower gas density within halos. It  
confirms that the new SPH implementation
substantially reduces over-cooling of gas.
In the simulation carried out with the conventional SPH
implementation, we found the opposite behaviour; the cold gas mass
fraction gets progressively higher at lower redshift.
This (mis-)behaviour agrees with the experiments of Springel \&
Hernquist (2001)
who showed that an SPH formulation with
geometric symmetrisation can substantially overestimate
the cooling rates in the centres of halos, leading
to the accumulation of too much cold gas.
The cold gas fractions
are near 80\% for most halos  
at $z=0$ in our simulation using this scheme, much larger than the
values plotted in Figure~\ref{fig_cool3}. Even for the main
cluster halo itself, the cold gas fraction is overestimated by
a factor of 2. 

Although the small scatter and the systematic trends in cold
gas fraction seen above a halo mass of $10^{12} h^{-1}M_{\odot}$ 
in Figure~\ref{fig_cool3} are in good agreement
with simple semi-analytic models (e.g. White \& Frenk 1991 and
Section 5 below), the increase in scatter below $10^{12}
h^{-1}M_{\odot}$ is unexpected. This is a consequence of resolution
problems in the SPH simulation. When a halo contains relatively few
gas particles, the smoothing inherent in the SPH technique reduces the
estimated gas density to the point where the cooling rate is seriously
underestimated. As is clearly visible in the figure, the effect is 
worse at $z=0$ than at higher redshifts because of the lower mean
cooling rates at later times. We will return to the effective
resolution limit implied by this problem in Section 5.

At redshift $z=1$ and particularly at $z=2$, another interesting
effect is visible at small masses in Figure~\ref{fig_cool3}. As the
typical cold gas fraction approaches 100\%  the cold gas mass within
halos rises systematically above the nominal maximum given by the
halo mass times the global baryon fraction. At the lowest masses
this effect is about a factor of two.  Although this may, in part,
be due to SPH artifacts in the strong cooling and small particle number
limit, much of the effect is undoubtedly real. When cooling is
efficient shocks are unable to raise infalling gas to the virial
temperature of a halo, no significant hot gas atmosphere is formed and
the accretion shock moves in close to the halo centre. 
The total gas mass within the nominal halo radius then increases since
pressure effects no longer impede infall at larger radii.
We note in passing that this effect is not correctly included in
the semi-analytic model we discuss below, which assumes that the
maximum mass which can cool in any halo is $f_{\rm baryon}M_{\rm halo}$.

We close this section by mentioning a different cooling 
problem. As noted above, it has been known for more than twenty years
that feedback of some kind is
needed in hierarchical cosmogonies to prevent excessive
cooling and star-formation at early times. 
Katz \& White (1993; see also Evrard, Summers \& Davis 1995;
Suginohara \& Ostriker 1998) demonstrated that such over-cooling 
occurs in hydrodynamic simulations of CDM universes
provided these are able to resolve the relevant early phases
of evolution. Subsequent work by Lewis\etal (2000) and Pearce\etal
(2001) confirmed that the cold gas fraction indeed
increases with increasing resolution. In our simulation, the cold gas
fraction of the main cluster is about 23\% at the present epoch (see the
rightmost point in
Figure~\ref{fig_cool3}). Although this value is substantially smaller
than the 55\% which we found in the simulation with the conventional SPH
implementation, it 
still appears too high to be compatible with observational
constraints on the fraction of baryons locked up in stars in rich
clusters (Balogh\etal 2001). This again confirms 
that efficient feedback mechanisms are 
required to regulate cooling and star formation in realistic 
models. Nevertheless, since our aim is to compare the treatment of
gas cooling in SPH and SA simulations, this `cooling catastrophe' 
is not directly relevant to our analysis.

\begin{figure}
\centering
\epsfxsize=\hsize\epsffile{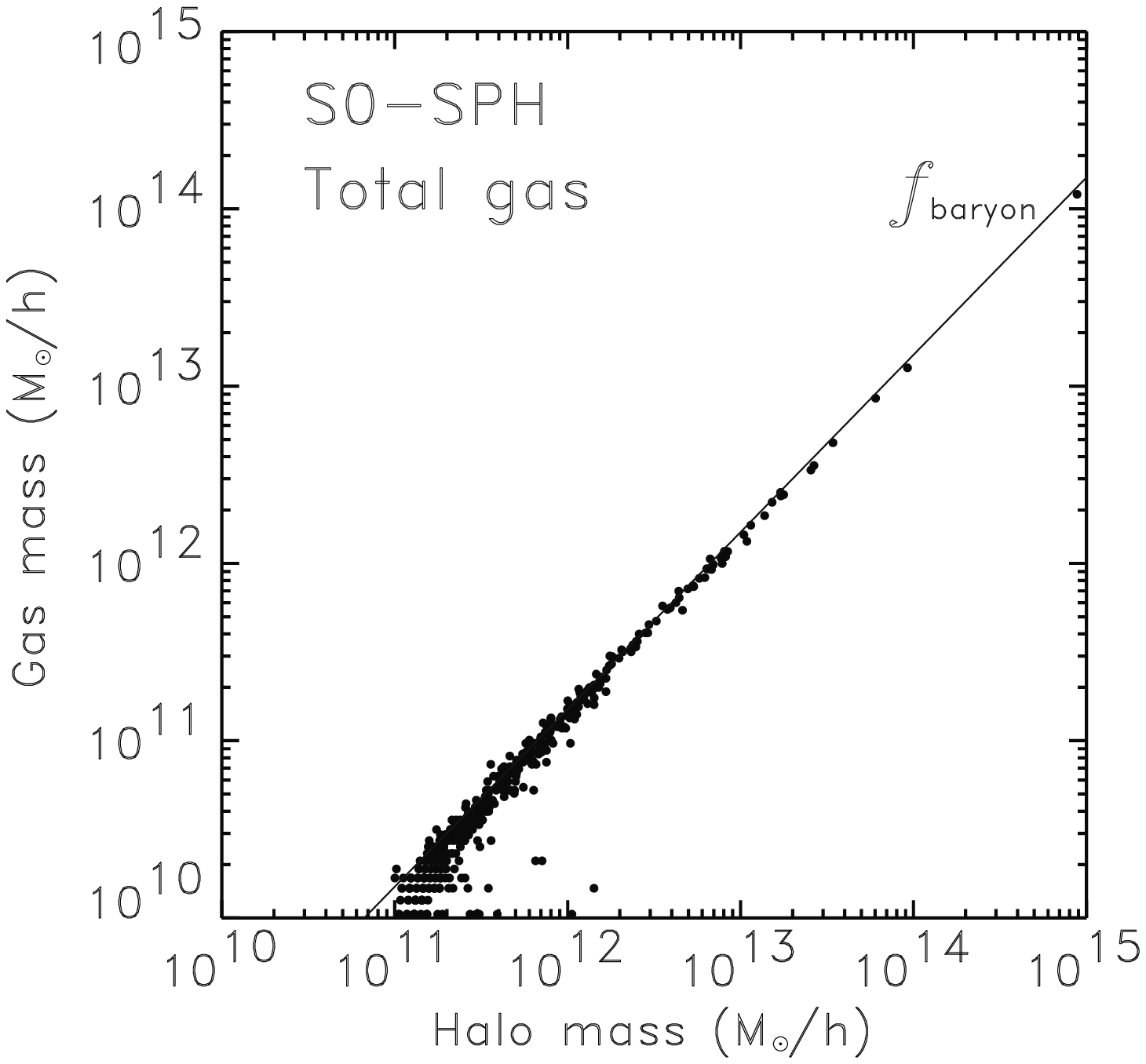}
\vspace{2mm}
\hspace*{3mm}\epsfxsize=0.95\hsize\epsffile{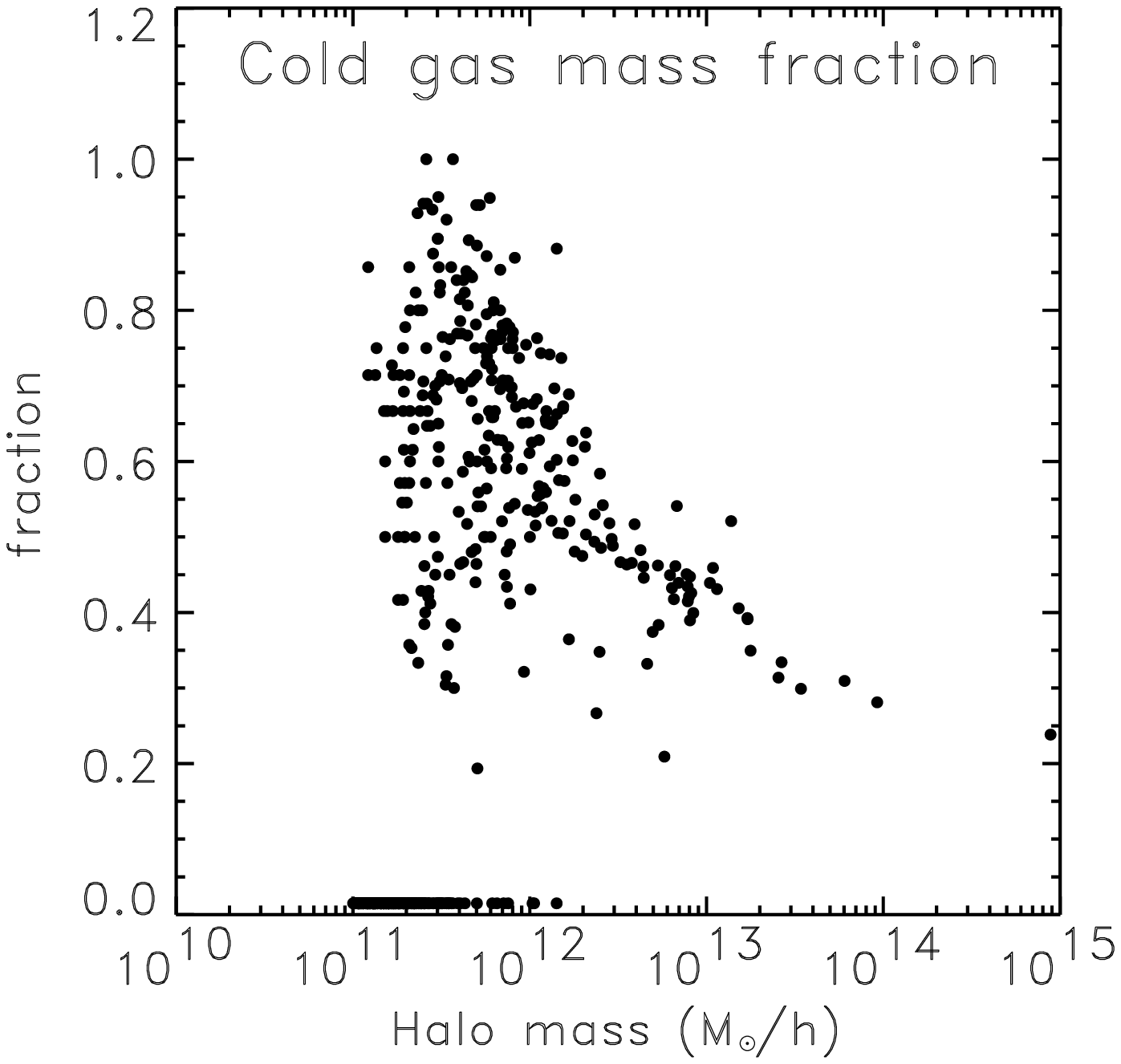}
\caption[Gas mass fractions in the S0 simulations.]
{We plot the total gas mass (top) 
and the cold gas fraction (bottom) against halo mass
at $z=0$. 
The solid line in the top panel indicates
the global baryon fraction. If a dark halo is empty of cold gas,
we assign a very small value instead of zero to the corresponding
fraction, so that those halos also appear in the figure.}
\label{fig_cool3}
\end{figure}

\setcounter{figure}{2}
\begin{figure}
\epsfxsize=\hsize\epsffile{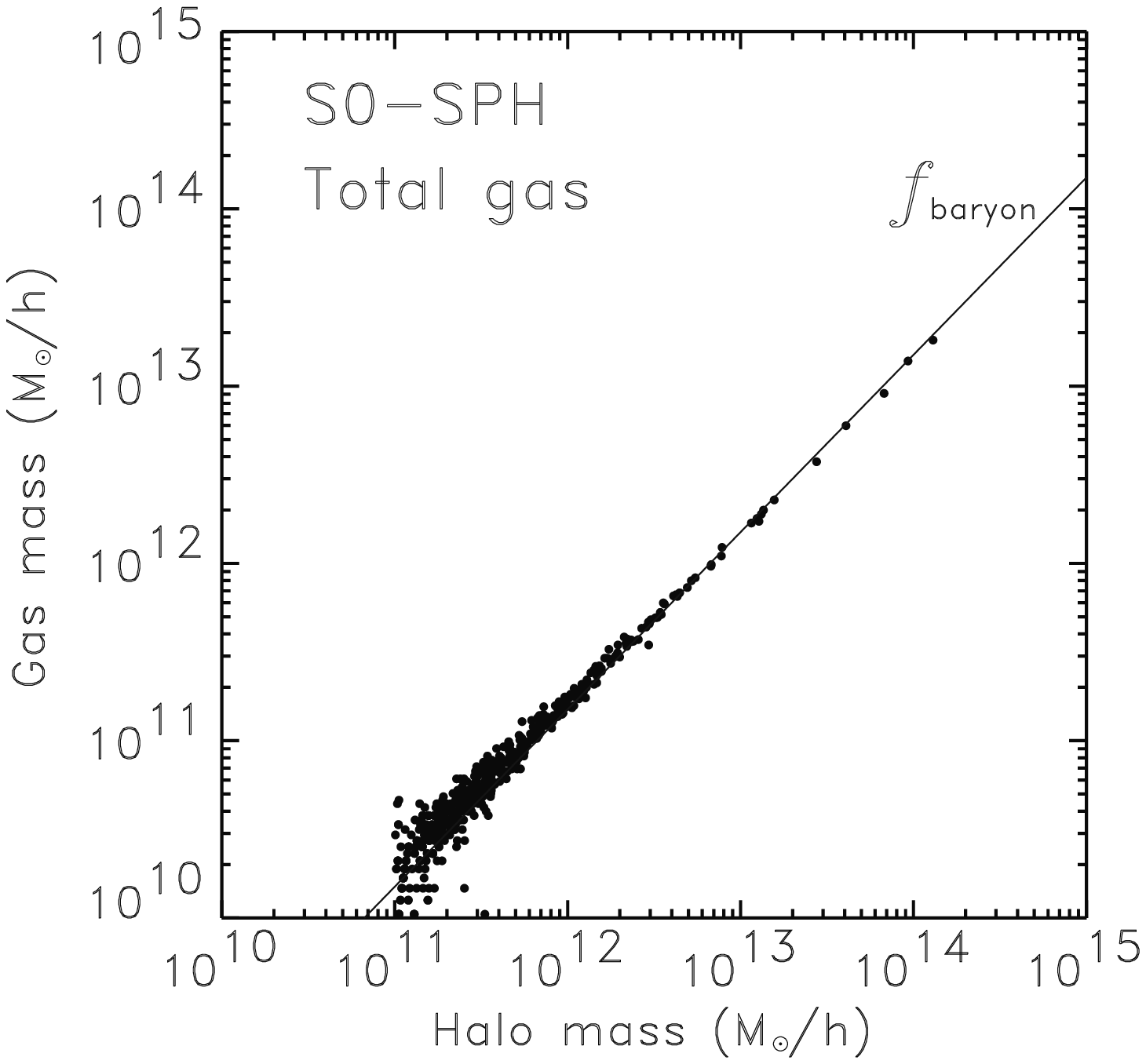}
\vspace{1mm}
\hspace*{4mm}\epsfxsize=0.95\hsize\epsffile{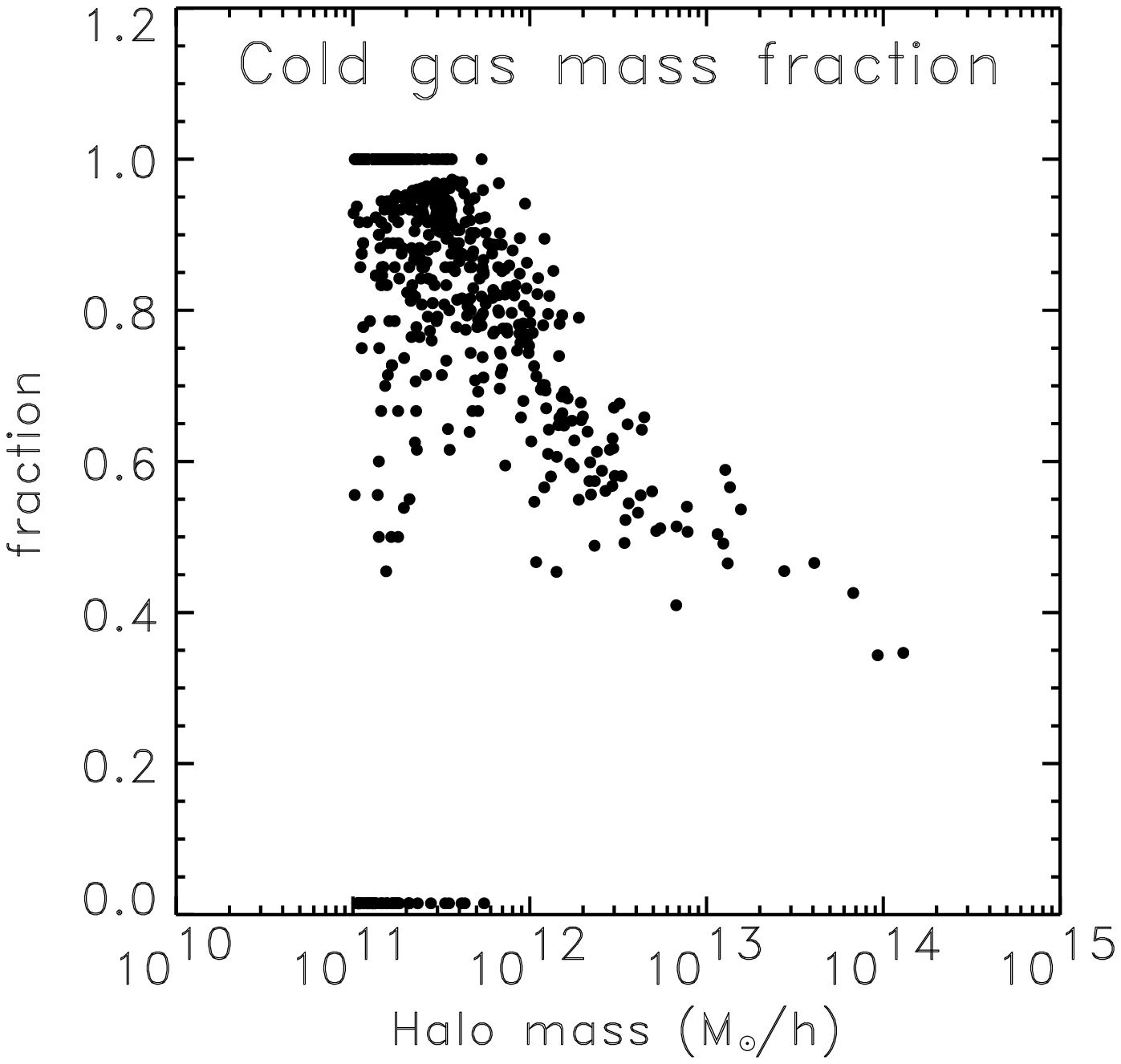}
\caption
{{\it -continued. } $z=1$.}
\end{figure}

\setcounter{figure}{2}
\begin{figure}
\centering
\epsfxsize=\hsize\epsffile{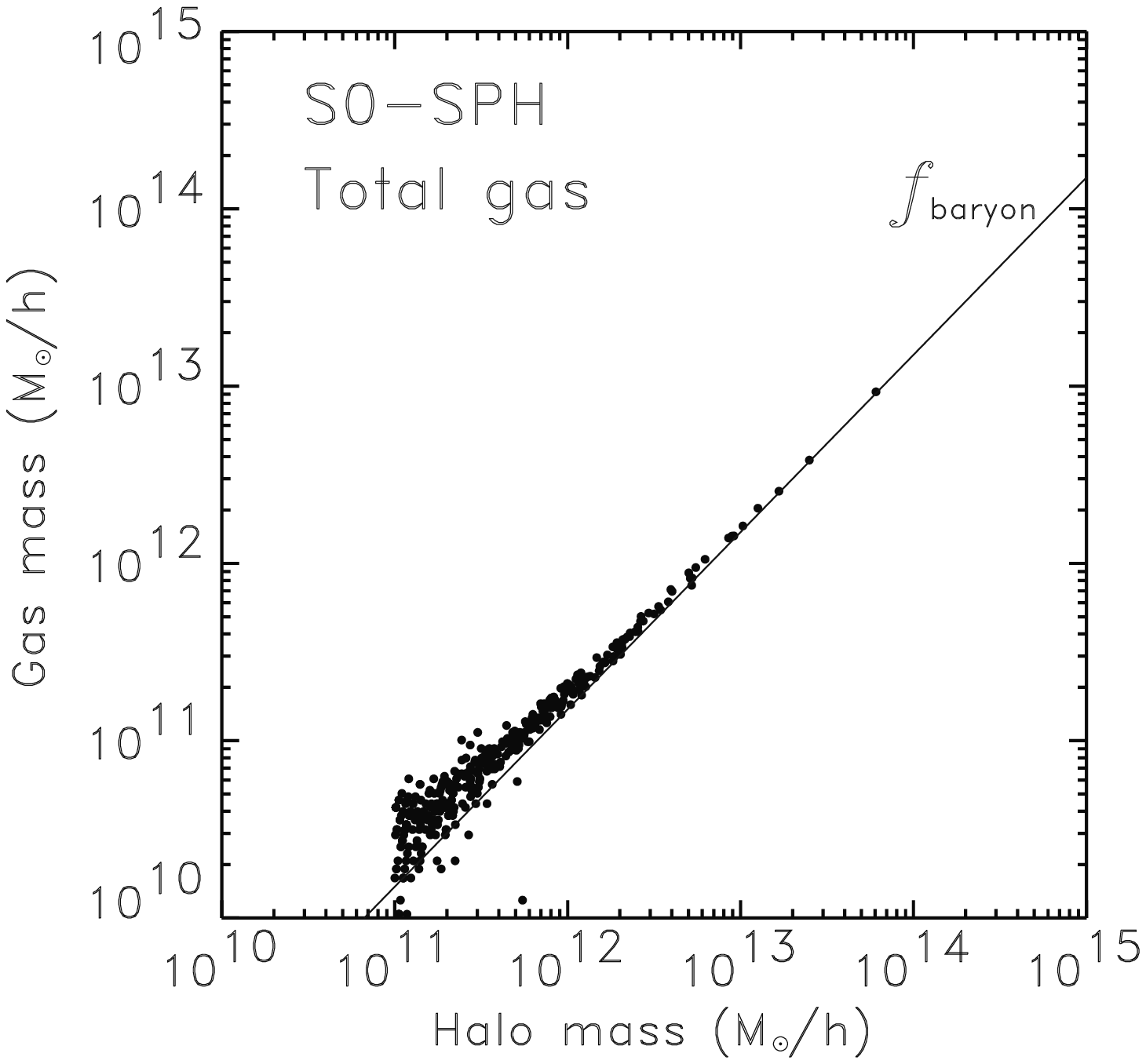}
\vspace{2mm}
\hspace*{3mm}\epsfxsize=0.95\hsize\epsffile{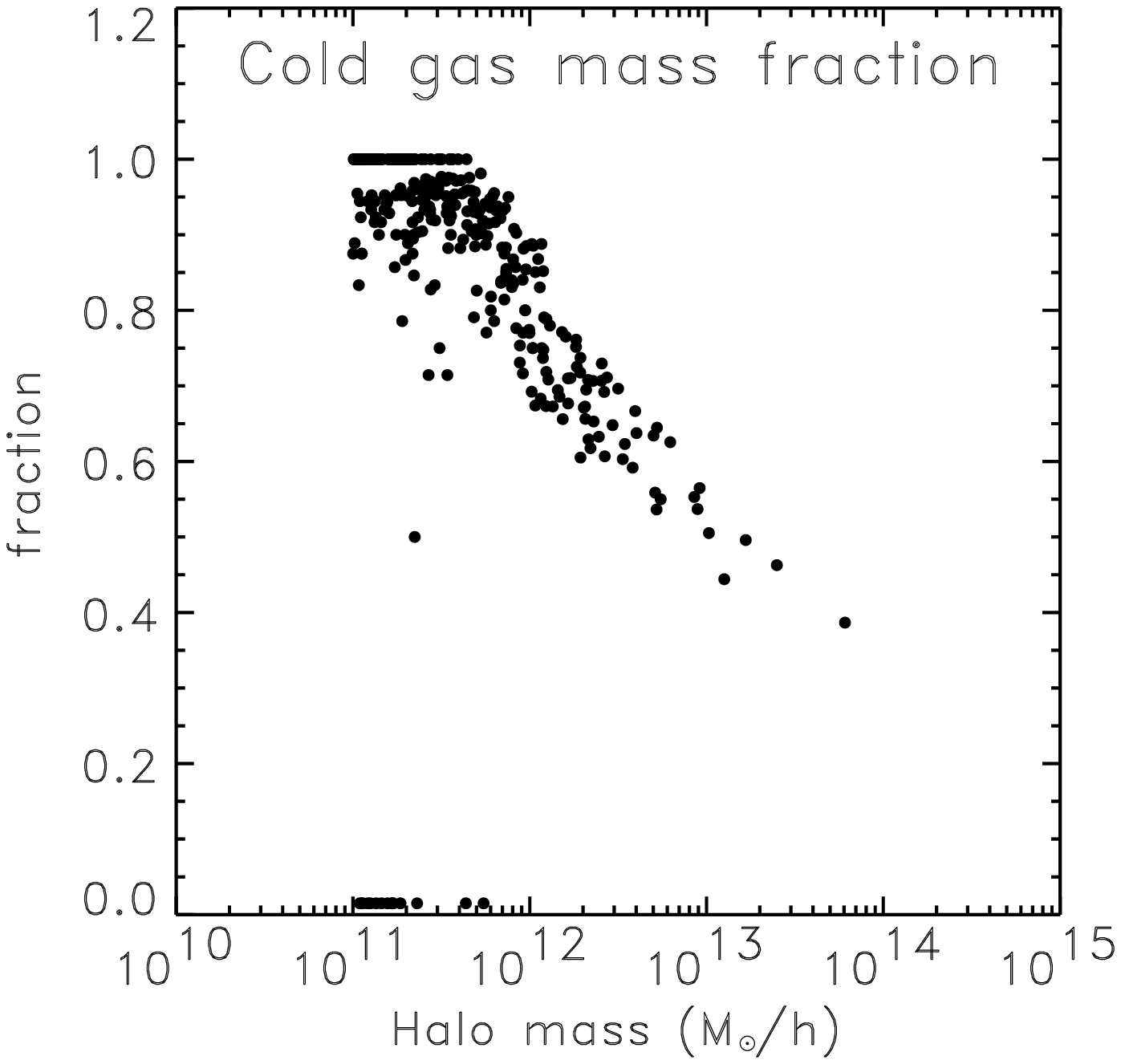}
\caption
{{\it -continued.} $z=2$.}
\end{figure}

\subsection{`Galaxies' in the SPH simulations}
In our two versions of the S0-SPH simulation, we define cold, dense 
clumps of gas as `galaxies'. We locate these objects by running a 
FOF group finder over the gas
particles with a very small linking parameter, $b=0.016$.
We note that the choice of linking parameter is not
particularly critical; adopting a slightly smaller or larger value
makes only a minor difference to the number of galaxies and their
masses. We also checked that for our chosen value most cold, dense 
particles in the more massive halos are linked into galaxies, and
multiple galaxies within a single halo are successfully
separated into discrete objects. In the following we require
a clump to contain ten or more gas particles to be considered
as a galaxy, corresponding to a 
minimum mass of $2.1\times 10^{10} h^{-1}M_{\odot}$. 
We refer to
such objects as SPH galaxies, in contrast to the SA galaxies 
defined using the semi-analytic model. The mass of an SPH galaxy
is simply the mass of its grouped particles.
We note that it may not be appropriate to regard $N_{\rm min}$ SPH particles 
as a distinct group, where $N_{\rm min} \le N_{\rm NGB}$, 
because of the SPH smoothing with $N_{\rm NGB}$ particles. 
Nevertheless we include such small groups in our SPH galaxy catalogue,
because, at early epochs, many of them are progenitor 
galaxies with which we can keep track of the formation history in detail.
We give an extensive discussion on the resolution of our SPH 
simulation in section 5.

\section{Semi-analytic simulations of galaxy formation}
In the galaxy formation models of KCDW 
and of SWTK, outputs of high resolution N-body 
simulations are used to define populations of dark halos and to 
follow their formation and evolution. Although the merging of halos 
is followed explicitly by these simulations (as is also the evolution
of halo substructure in SWTK), all
aspects of the evolution of the baryonic component are included
using simple, physically motivated, recipes taken from earlier
semi-analytic models. In this paper we follow the procedures of
KCDW, simplified to exclude processes other than
gas cooling and `galaxy' merging. This eliminates the free
parameters normally adjusted in SA models, since these
are associated with star formation and feedback. The standard
cooling model contains no free parameters (see below) and the
merger model of KCDW implies merger rates
in good agreement with those seen in simulations where
merging can be followed explicitly (SWKT).

For each dark halo in each simulation output, we assign the
following related halo properties: the virial radius $R_{\rm vir}$,
the virial mass $M_{\rm vir}$, and the circular velocity
$V_{\rm vir}^{2} = G M_{\rm vir}/R_{\rm vir}$. These halos
are linked to their progenitors in the previous output to build
up merger history trees within which the growth of galaxies
is followed. Notice that since we use the dark matter distribution
from the $N$-body/SPH simulation to set up these trees, there is an
exact correspondence between the halos and their merger histories in
our SA and SPH simulations. Any differences in the resulting
galaxy populations must therefore be produced purely by differences 
in how gas dynamics, cooling and galaxy merging are handled.
We now turn to the assumptions used to model gas cooling.
 
\subsection{Gas cooling}
Gas cooling is treated as in White \& Frenk (1991). We assume that
the gas in a dark halo is shock-heated to the virial temperature
and is initially distributed  like a singular isothermal sphere
with density profile $\rho_{\rm g}\propto r^{-2}$.
We define the local cooling time of the gas,
$t_{\rm cool}$, as the ratio of the thermal energy density
of the gas to the cooling rate per unit volume:
\begin{equation}
t_{\rm cool} (r)=\frac{3}{2}\frac{kT \rho_{\rm g}}
{\bar{\mu}m_{\rm p} n_{\rm e}^{2}(r)\Lambda (T,Z)},
\end{equation}
where $\bar{\mu}$ is the mean molecular weight, $m_{\rm p}$
is the proton mass, $n_{\rm e}$ is the electron number density,
$T$ is the gas temperature, which we approximate with the virial
temperature of the halo $T=35.9 (V_{\rm vir}/{\rm km s}^{-1})^{2}$ K,
and $\Lambda (T,Z)$ is the cooling rate. We employ the tabulated cooling
functions of Sutherland \& Dopita (1993).
 
We define a cooling radius $r_{\rm cool}$ such that
the cooling time at this radius is equal to the time for
which the halo has been able to cool quasi-statically.
We approximate this time with the halo's dynamical time $R_{\rm
vir}/V_{\rm vir}$. The cooling rate for the gas within this cooling
radius is then given by
\begin{equation}
\frac{{\rm d} M_{\rm cool} }{{\rm d}t}
=4\pi \rho_{\rm g}(r_{\rm cool}) r^2_{\rm cool}
\frac{{\rm d} r_{\rm cool} }{{\rm d}t}.
\end{equation}
In reality, cooling rates depend strongly on the metallicity of the
gas, but for simplicity (and consistency) we use zero metallicity 
cooling functions in both our SA and our SPH models.

We note that this cooling model is extremely simplified and corresponds quite
poorly to the detailed dynamics seen in hydrodynamic simulations of
the evolution of galaxy-sized objects in a CDM universe. The good
agreement which we find below with the gas accumulation rates in our best
SPH simulation is thus quite surprising.
 
\subsection{The stripped-down semi-analytic model}
Our SPH simulations do not include star formation or feedback
processes. In order to make a direct comparison, we remove these
processes also from our semi-analytic simulation.
In this stripped-down model, star formation and feedback are
switched off, and no cooling cut-off is implemented in massive
halos (c.f. KCDW). 
As in the SPH simulations, galaxies in our stripped-down SA
simulation do not possess properties such as luminosity
or morphological type. They are defined purely by their mass of cold
gas, their positions and velocities, and their formation histories.
There are only two populations of galaxies in this model. Each dark
halo carries exactly one {\it central} galaxy which is identified
with the most-bound particle of the halo.
Many halos have also one or more {\it satellite} galaxies.
Satellites are galaxies that were central galaxies
in earlier outputs, but whose host halos merged at some time
into the larger halo where they now reside.
Satellites are identified with the particle to which they were
attached when they were last a central galaxy.
They are assumed to merge with the central galaxy of their
new halo on a dynamical friction timescale (see KCDW). In
this implementation, only the central galaxy accretes
new gas that cools within the halo; hence the mass of a satellite
galaxy stays constant.

\section{The resolution of the SA and SPH simulations}

In this section we discuss the effective resolution limits
of our two simulation techniques. The principal goal of our
paper is to compare the predictions of these techniques for the
cooling of gas onto protogalaxies, but before this is possible we
must establish the range of protogalaxy masses for which
the predictions of each technique are independent of the
resolution parameters -- particle mass, force softening,
SPH smoothing, timestep -- employed in the corresponding simulation.
With this in mind, we define the effective resolution of a simulation
to be the minimum galaxy mass for which the simulated galaxy 
population has identical properties to that in a much
higher resolution simulation of the same system {\it carried out 
with the same technique}. Notice that with this definition there
is no guarantee that the galaxy populations produced by simulations
using our two techniques will agree above the larger of their
two resolution limits, although we find below that they do.
Notice also that for the physical processes included here -- cooling,
but no star formation or feedback -- there is reason to worry that our
definition may fail; as the resolution of a simulation is increased
the masses of {\it all} galaxies might increase systematically towards
their maximal possible values, corresponding to the total baryon 
content of their individual halos. We find below that a `cooling
catastrophe' of this type does not, in fact, occur.

\begin{figure}
\centering
\epsfxsize=\hsize\epsffile{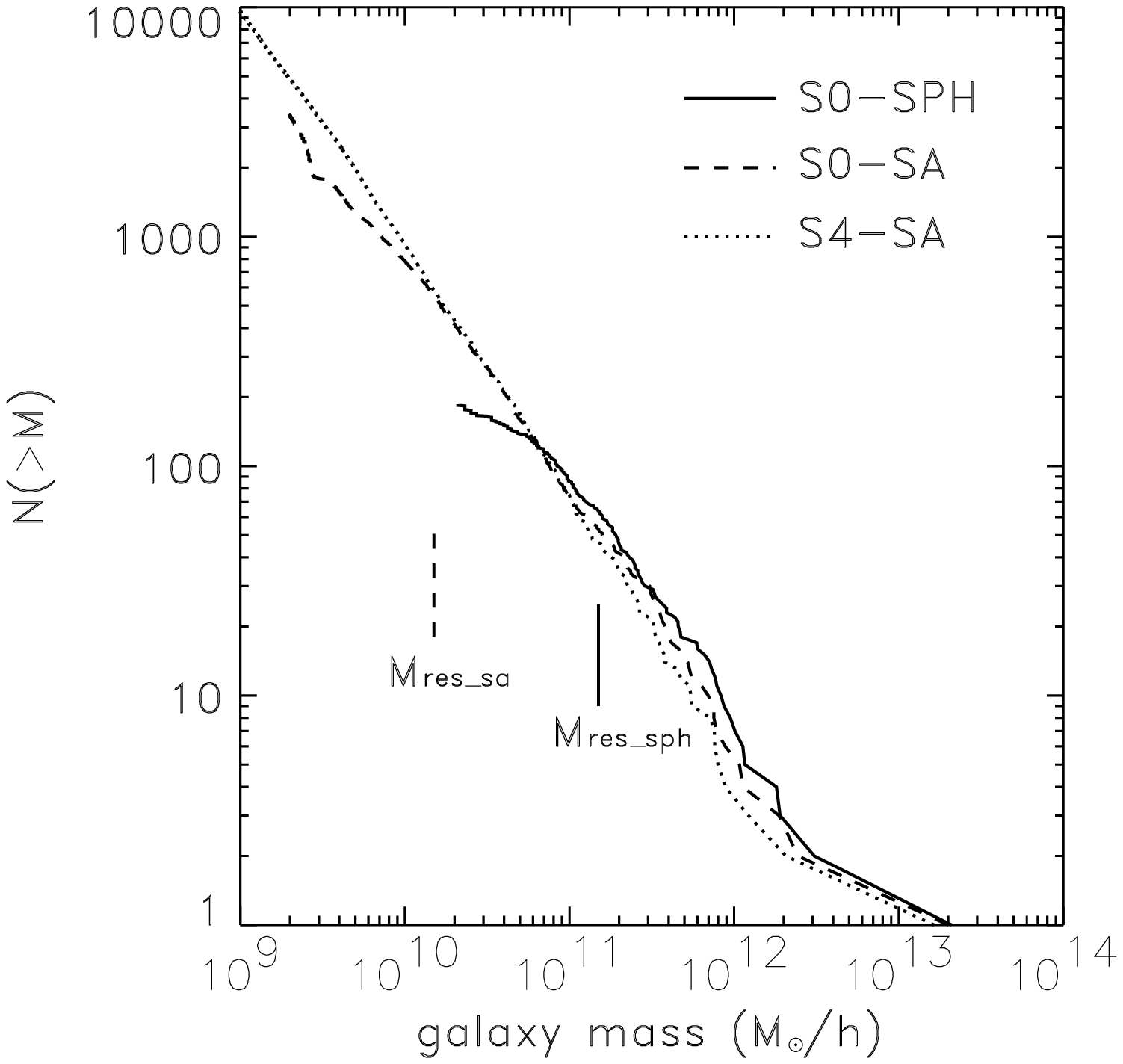}
\caption[The galaxy mass functions.]
{The cumulative mass functions of the `galaxies' (i.e. the cold clumps of 
dense gas) in the semi-analytic and SPH simulations.
We indicate the effective mass resolutions discussed in the text
using vertical bars. The galaxy mass function for the S4-cluster
(the highest resolution simulation of the same cluster in Springel
\etal 2001b) is plotted for comparison. The mass resolution of this 
simulation is 250 times better than that of the SA simulation studied 
in this paper.}
\label{massfunc}
\end{figure}

We begin by studying the resolution limit of our SA simulation
technique. This is relatively straightforward because SWTK 
have already carried out a series of higher resolution simulations
of the physical system studied here. For their highest resolution
model, S4, the mass resolution is about 250 times higher and the
spatial resolution about 50 times higher than in our S0-SA simulation. 
We have run our `stripped-down' SA modelling on SWTK's S4 simulation 
to derive a galaxy population which can be compared
directly with those of this paper. Figure~\ref{massfunc} compares
the cumulative mass functions of the two SA simulations. Above a mass
of about $1.5 \times 10^{10} h^{-1}M_{\odot}$ the agreement is good
with the masses of galaxies in the high resolution simulation lying
slightly below those of galaxies in S0-SA. At lower mass the
mass function of S0-SA shows a clear change in slope. This is easily
understood. Our halo merger trees keep track of all halos with 10
or more particles. The maximum mass of a galaxy which could form
in an unresolved halo is thus given by the global baryon fraction
times the mass of a 9 particle halo. This is
$1.9 \times 10^{10} h^{-1}M_{\odot}$, slightly larger than the mass at
which we see the change in slope. The convergence tests of
SWTK indicate similar resolution limits for their
full semi-analytic model.

It is important to note the very steep mass functions in
Figure~\ref{massfunc} which are a consequence of the absence of
feedback. The total amount of gas cooled in the SA simulations is a
strong function of resolution, rising from 16\% in S0-SA to
25\% in S4-SA. (We have not attempted to include the effects of an
ionizing background in the SA modelling, which would reduce the
strength of this effect.) It turns out, however, that the additional
gas which cools off in small objects does not reduce the reservoir
of diffuse gas enough to substantially affect the masses of the larger
objects which form later. Similarly merging is not efficient enough
for the increased supply of low-mass merger candidates to cause a 
substantial increase in the masses of larger systems. Thus, although
there is no formal convergence as resolution is increased, the
properties of the higher mass galaxy population vary very little.

It is more difficult to study the mass limit 
for our SPH model, because we cannot carry out a similar
convergence test to that we decribed above for the SA models.
Our decision to impose a 10 particle
lower limit on our SPH `galaxies' clearly imposes a lower bound
of $2 \times 10^{10} h^{-1}M_{\odot}$ on the resolution limit, but
Figure~\ref{fig_cool3} shows that the effective resolution limit is
substantially higher. Below a {\it halo} mass of about
$10^{12} h^{-1}M_{\odot}$ many halos contain substantially less
cold gas than expected and some contain none
at all. This is a consequence of the smoothing associated
with SPH density estimates which leads to an underestimate of the
cooling in systems where the number of gas particles is comparable to 
that required within the SPH smoothing kernel. We plot the $z=0$
cumulative galaxy mass function for the `entropy' version of 
our S0-SPH simulation in Figure~\ref{massfunc}. Above 
$1.5 \times 10^{11} h^{-1}M_{\odot}$ this function agrees 
well with that of our two SA simulations, except for a slight 
shift to higher mass. Below $1.5 \times 10^{11} h^{-1}M_{\odot}$
the mass function has a clear change in slope. 
We take this mass scale as a conservative estimate for
the mass resolution of our SPH simulation.
This effective
resolution limit in our S0-SPH simulation is about
an order of magnitude larger than the corresponding limit
for our S0-SA simulation. In light of this a robust comparison 
of SA- and SPH-galaxies can only be made at masses larger than
about $10^{11} h^{-1}M_{\odot}$.

\begin{figure}
\centering
\epsfxsize=0.8\hsize\epsffile{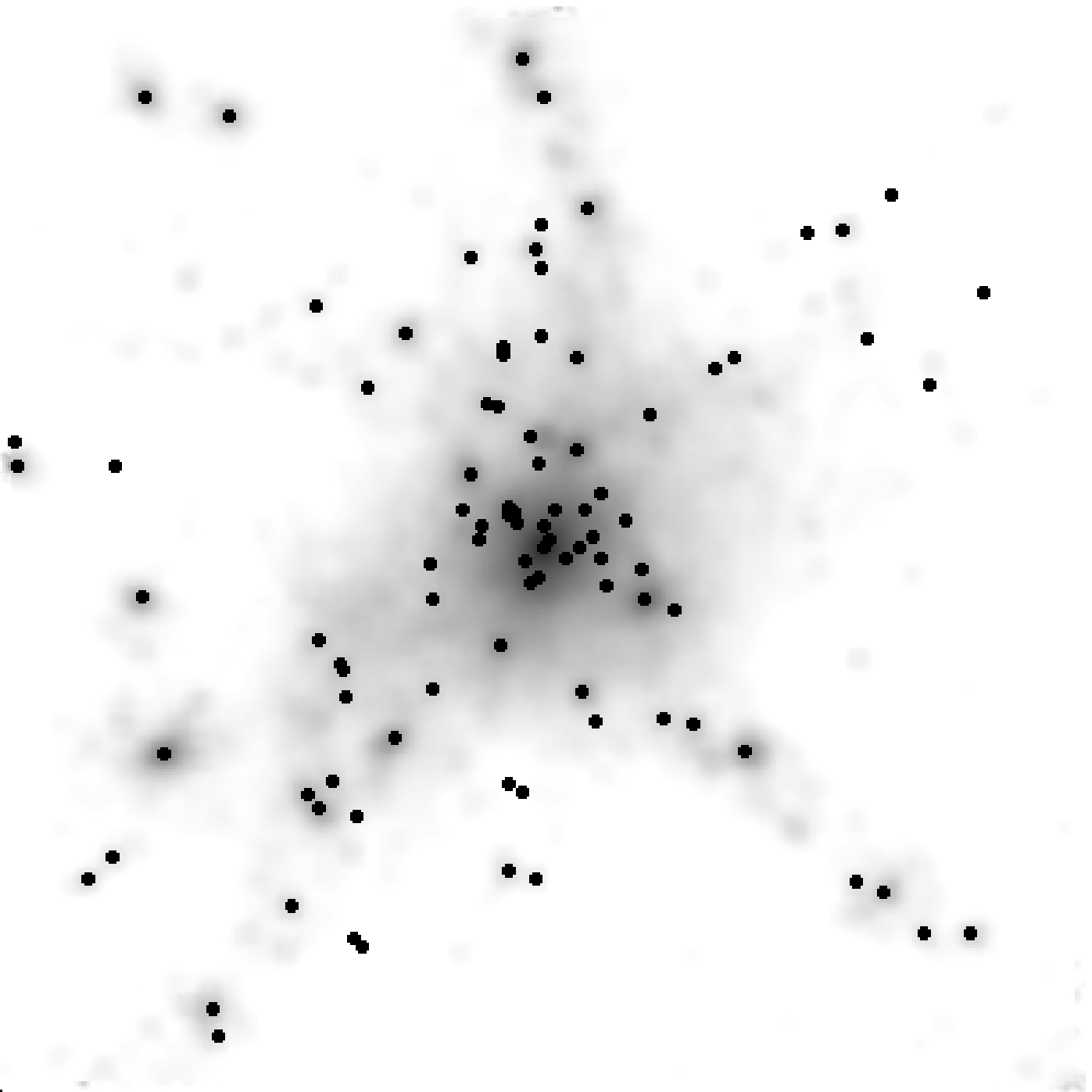}
\vspace{6mm}%
\epsfxsize=0.8\hsize\epsffile{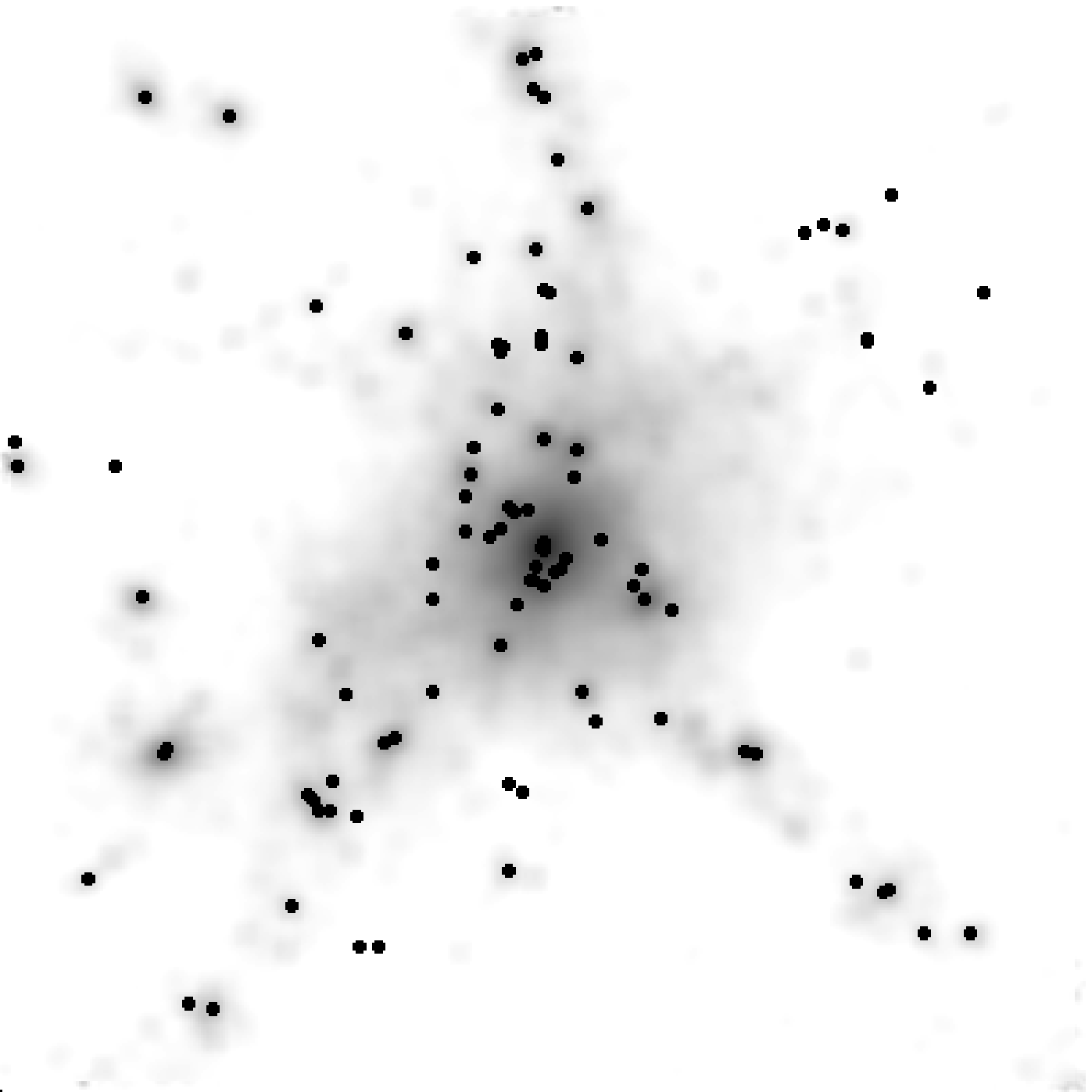}
\caption[Distribution of galaxies in the S0 simulations.]
{Distribution of SA-galaxies (top) and SPH-galaxies (bottom)
at $z=0$. Only galaxies with mass larger than $6\times
10^{10} h^{-1}M_{\odot}$ are plotted in the figures.
The grey scale shows the projected density (gas and dark matter)
distribution.}
\label{fig_cool5}
\end{figure}

\section{Object-by-object comparison}
We now turn to a direct comparison of the properties of the
individual objects which form in our SA and SPH simulations.
The dark matter distributions of the two simulations are
identical at all times, by construction, and as a result it is
possible to identify individual galaxies in the two models almost without
ambiguity. We begin by comparing the masses of the central galaxies. 
Every halo with at least 10 particles is automatically assigned a
central galaxy in our semi-analytic model (see Section
4.2) whereas defining the central galaxy in the SPH simulation
is less clear. For simplicity, we define the most massive
SPH-galaxy within a halo to be its central galaxy, and we tag 
other galaxies as satellites. Note that this definition
corresponds closely to that used in the SA model for halos with
multiple galaxies (see KCDW).

In Figure~\ref{fig_cool5} we compare the distributions of SA- and 
SPH-galaxies more massive than $6\times 10^{10} h^{-1}M_{\odot}$
in a cubic region, 15$h^{-1}$Mpc on a side, centred on our cluster. 
The two distributions are very similar, and an almost one-to-one 
correspondence can be found. Most of these galaxies are the
central galaxies of their dark matter halos except in the main 
cluster itself where there are many satellites. Thus the galaxy
distribution largely reflects that of the centres of the
massive halos. Most differences arise when galaxies lie on 
opposite sides of our mass threshold in the two samples.
       
In Figure~\ref{fig_cool7} we plot SA mass against SPH mass for the 
central galaxies of all halos with 10 or more dark matter particles.
Halos with no SPH-galaxy are shown on the left of this diagram at
an arbitrary SPH mass. Above our SPH effective resolution limit 
($\sim 10^{11} h^{-1}M_{\odot}$) the central galaxy masses predicted
by our two techniques agree extremely well over more than two orders 
of magnitude in mass. Below this limit the masses of the SPH-galaxies
are systematically low and many halos contain no SPH-galaxy even
though the corresponding SA-galaxy mass is well above 
$2\times 10^{10} h^{-1}M_{\odot}$, the limit imposed by our 
requirement that an SPH-galaxy contain at least 10 particles.
We emphasize that the agreement above $10^{11} h^{-1}M_{\odot}$
limit is remarkable considering the crudity of the cooling model used
in the SA treatment and the fact that our two implementations of SPH
lead to galaxy masses differing systematically by a factor of two. 
(In this section we only show results from the `entropy' implementation.)

 It is also interesting to see the {\it evolution}
of the masses in two models. In Figure~\ref{massevolution} we compare
the masses of the three most massive galaxies at $z=0$ and their progenitors
at earlier epochs.  It is clearly seen that even the time evolution is 
remarkably similar in the two models.
Small deviations from the perfect linear relation at output instants 
are due to time offset of satellite merging onto the central galaxy,
because of the simplified prescription of satellite merging 
in our SA model.

\begin{figure}
\centering
\epsfxsize=\hsize\epsffile{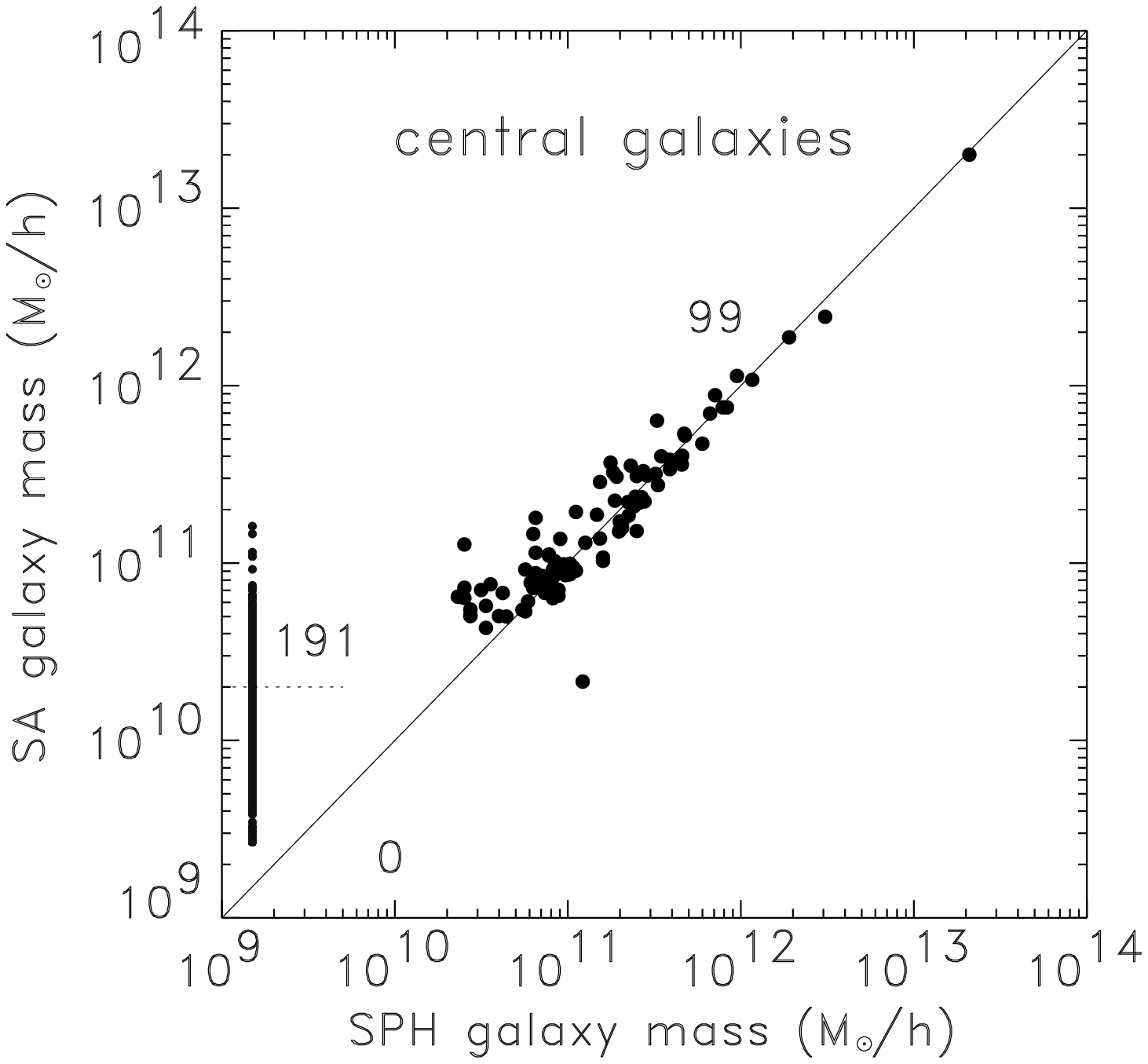}
\epsfxsize=\hsize\epsffile{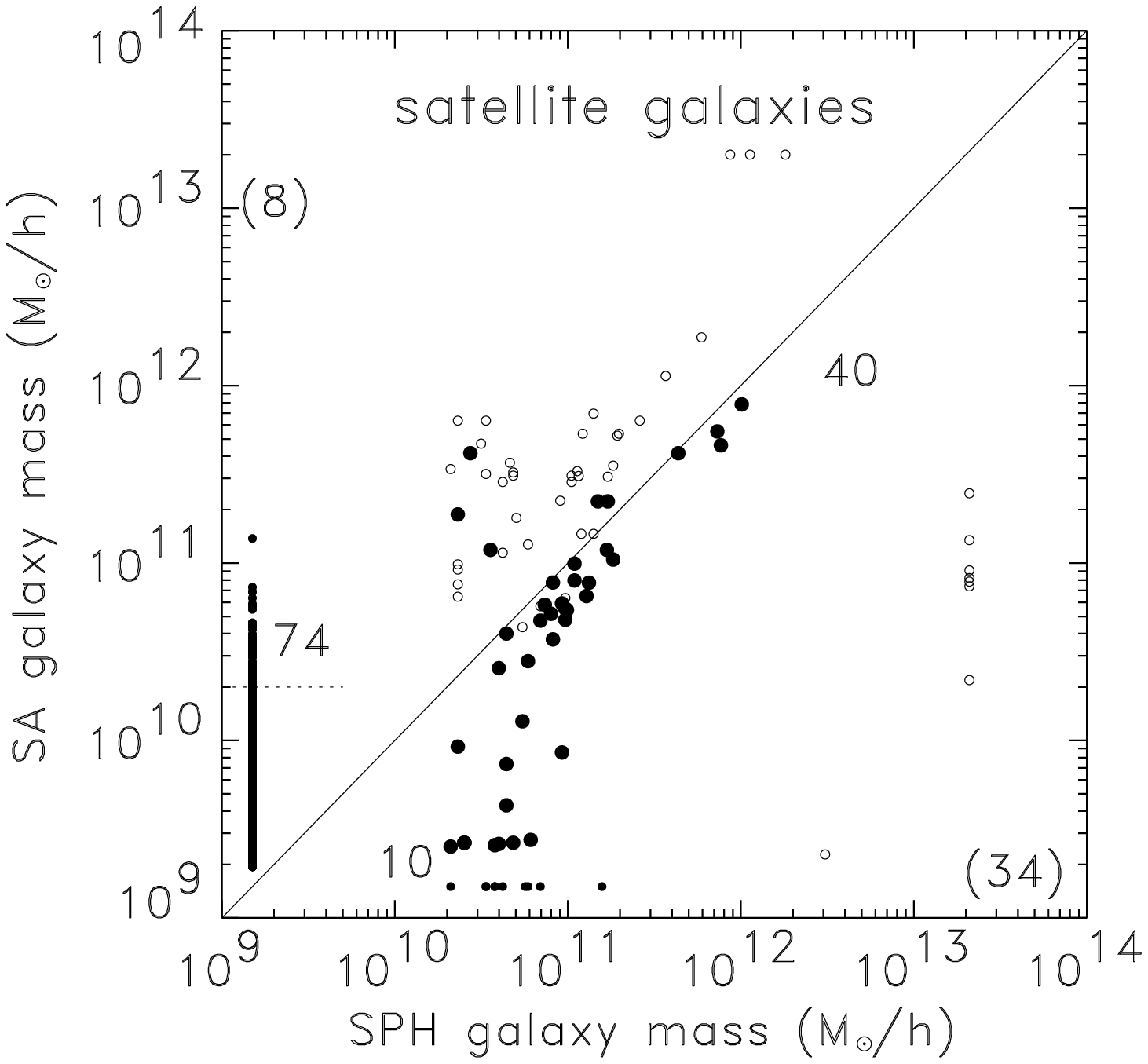}
\caption[Galaxy mass comparison.]
{The SA-galaxy mass is plotted against the SPH-galaxy mass
for central galaxies (top) and for satellite galaxies (bottom).
If a dark halo hosts a SA(SPH) central galaxy but no SPH(SA) galaxy,
we assign a very small value instead of zero as the 
corresponding SPH(SA) galaxy mass. These galaxies then
also appear in the figure.
The number of such galaxies is given beside
the points at very low mass. For satellite galaxies, 
we plot the satellite - satellite matching cases
as filled circles.
The numbers in parentheses are the number of SPH-satellites
which are matched to SA-centrals (right lower corner)
and of SA-satellites which are matched to SPH-centrals (left upper
corner). These galaxies are plotted as open circles.}
\label{fig_cool7}
\end{figure}

\setcounter{figure}{5}
\begin{figure}
\centering
\epsfxsize=\hsize\epsffile{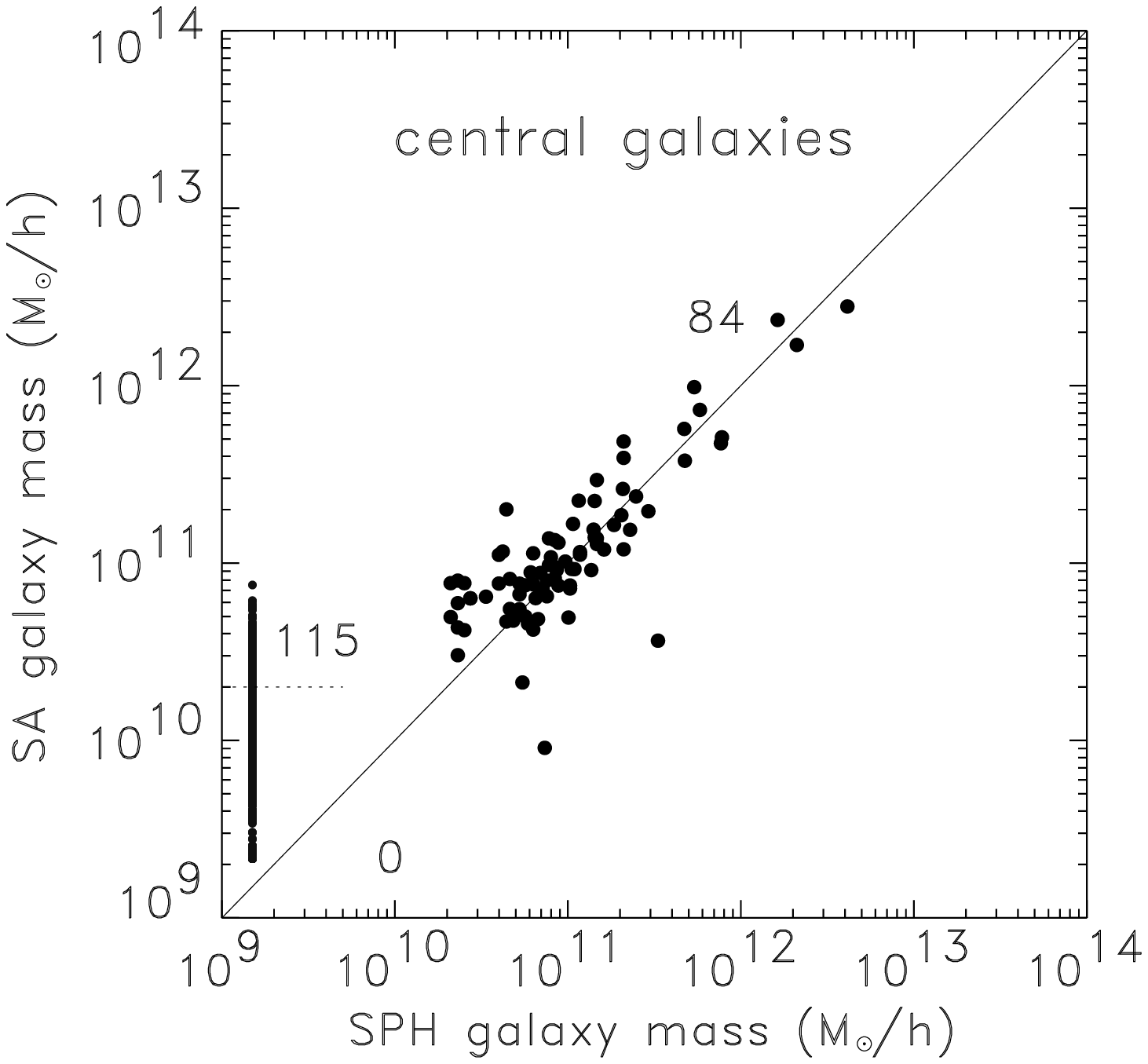}
\epsfxsize=\hsize\epsffile{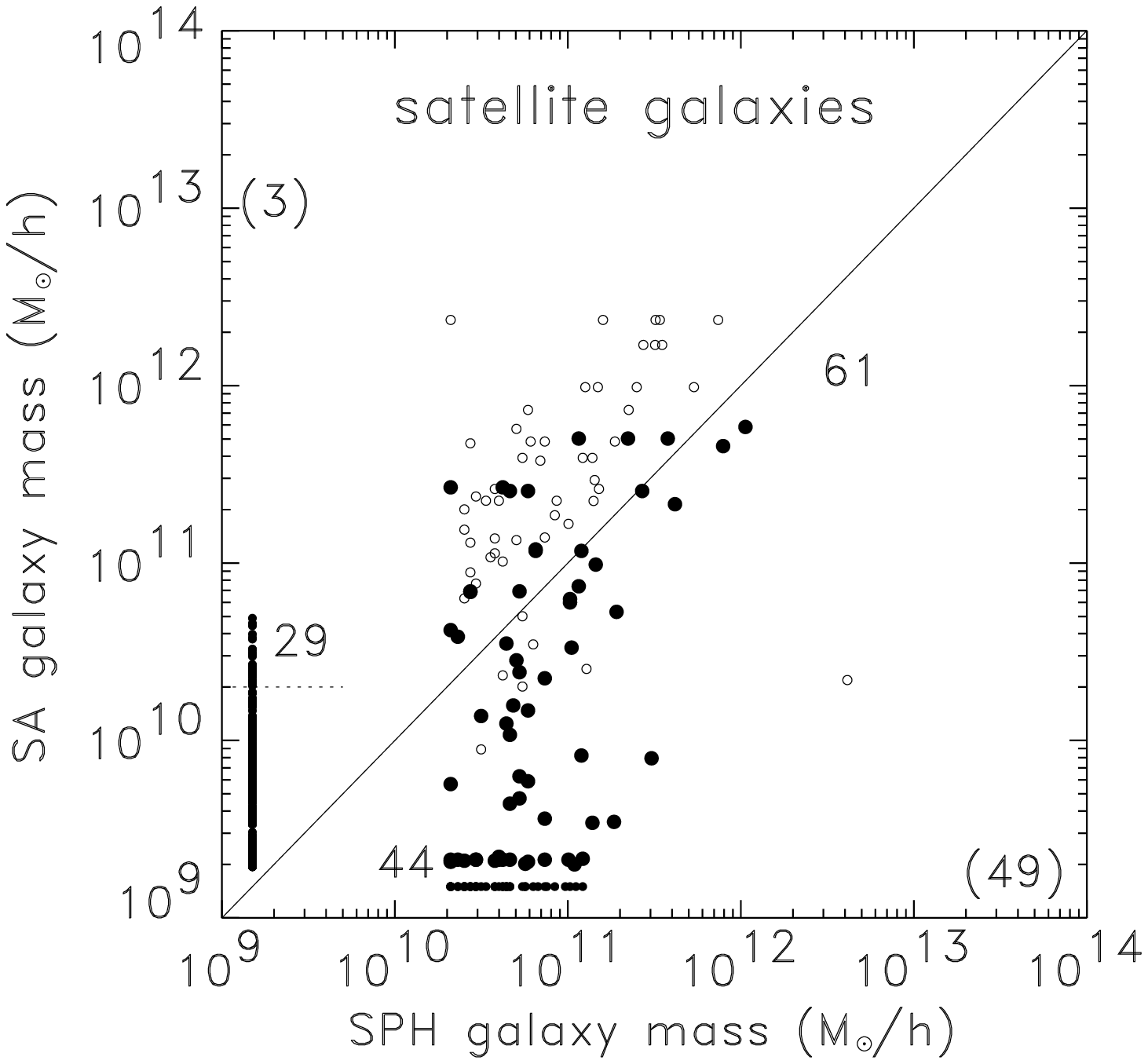}
\caption{-{\it continued.} $z=1$.}
\end{figure}

\setcounter{figure}{5}
\begin{figure}
\centering
\epsfxsize=\hsize\epsffile{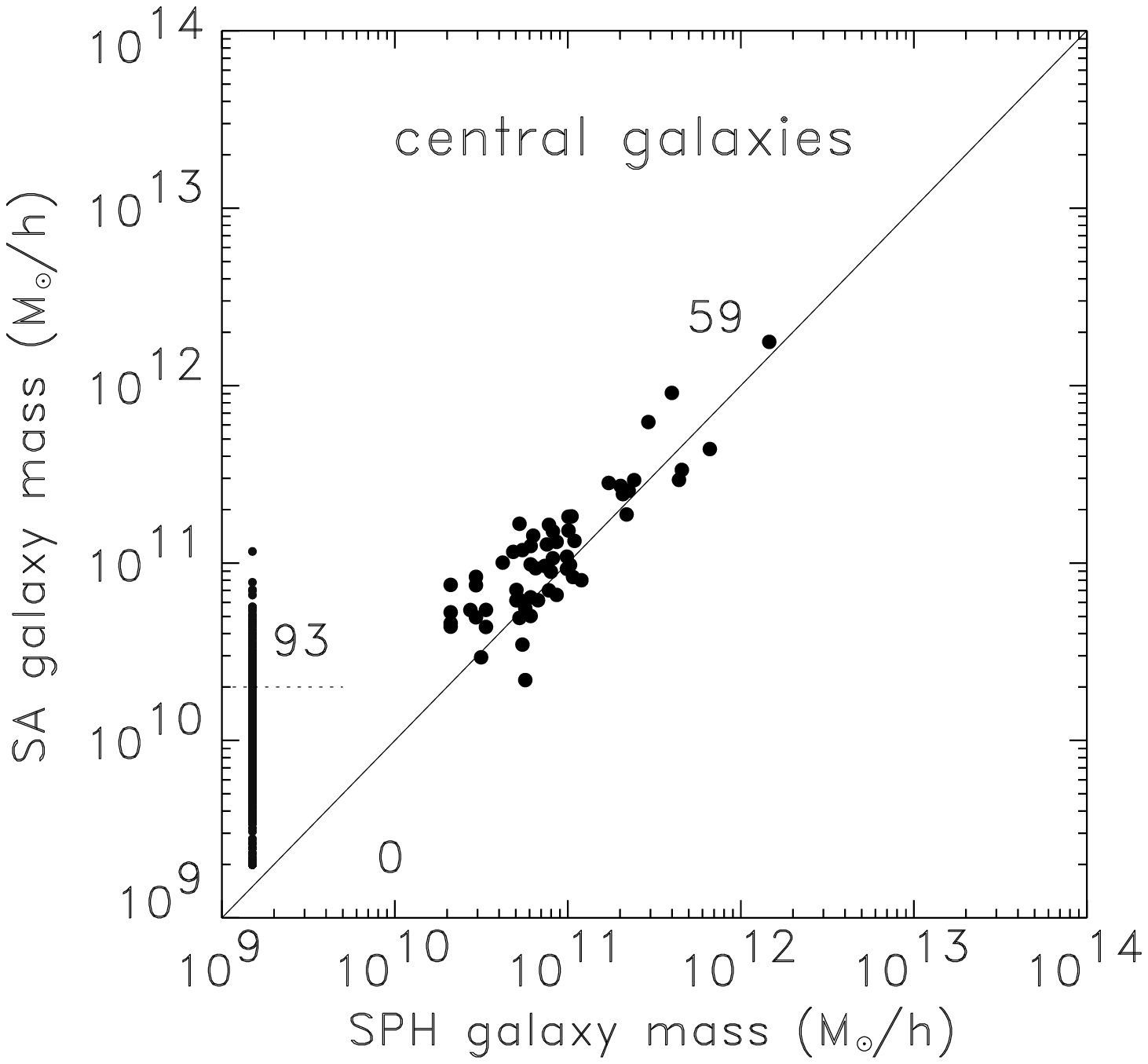}
\epsfxsize=\hsize\epsffile{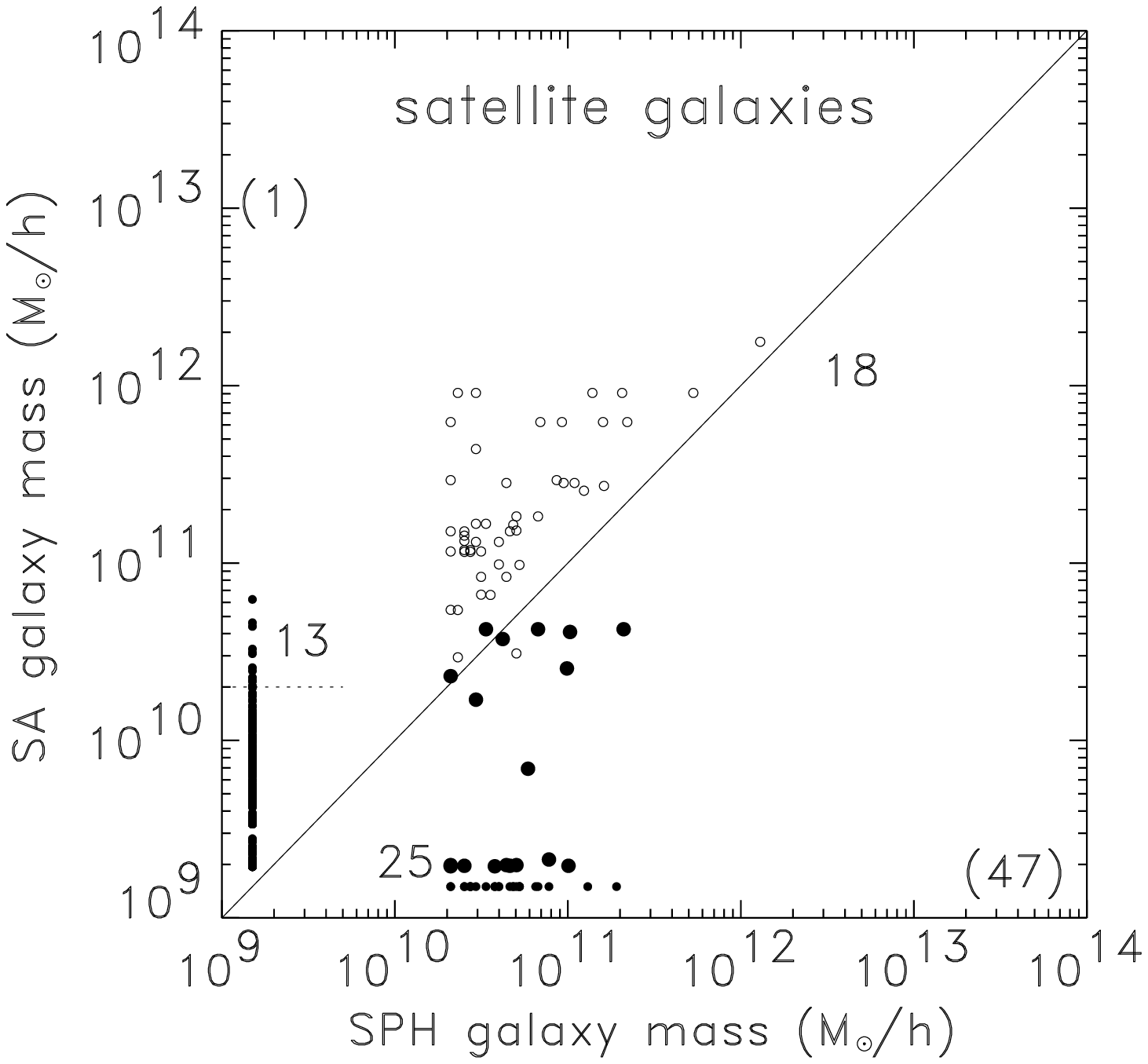}
\caption{-{\it continued.} $z=2$.}    
\end{figure}

A similar comparison for satellite galaxy masses is shown
in the bottom panels in Figure~\ref{fig_cool7}. In this case 
matching SA- and SPH-galaxies is much more complicated than 
for central galaxies. Once an object becomes a satellite its 
position will not be the same in the two kinds of simulation, 
and indeed it may merge with the central galaxy in one of 
the simulations but not in the other. We use the histories 
of the satellite galaxies in the two simulations in order 
to carry out this matching. Given a satellite galaxy identified 
in an output of the semi-analytic simulation, say at redshift 
$z_{1}$, we track it back through earlier outputs until we find 
the time when it was last a central galaxy. Let us denote this
time $z_{2}$. (Recall that, in our semi-analytic scheme every 
satellite galaxy was a central galaxy at some earlier time.)
We then search in the SPH simulation at $z_2$ and identify the
SA progenitor of our satellite with the most massive SPH galaxy 
in its halo. In some cases the halo may contain no SPH galaxy
and we then consider our original SA satellite to be
unmatched. Otherwise we locate the particles of the matched
SPH galaxy in the SPH simulation at $z_1$ and consider
the SPH galaxy which contains the majority of them to be the
counterpart of the original SA satellite. This counterpart
may be an SPH satellite, the SPH central galaxy, or (rarely)
an `orphan' galaxy outside any halo. Occasionally it
can fall below the 10 particle threshold to be considered a 
{\it bona fide} SPH galaxy and we then again consider the
original SA satellite to be unmatched.

We also carry out the reverse procedure. Starting
from an SPH satellite at $z_{1}$, we trace back
its progenitors (defined at each time to be
the SPH galaxy which donates the largest number of particles
to the galaxy at the subsequent output) until we find one which
was the most massive SPH galaxy in its halo. This one we then identify
with the SA central galaxy of the same halo. 
We take particular care to investigate SPH satellite galaxies
which appear to have no progenitors. 
In each output, some new SPH galaxies appear for the first time
as satellites; in the previous output, their member particles are 
already cold and are just condensing at the center of a halo. 
Due to discrete sampling of the output times, there is always a slight offset
between the true formation epoch of SPH galaxies and the time when they are
first identified. Their host halo can merge with a larger system
during this time. To take this into account, we tag the central SA
galaxy in any halo that hosts such a {\it forming} SPH galaxy as 
its counterpart, because it is clear that such pairs indeed
correspond closely. In this way we can partially compensate for
incompleteness in our SPH--SA galaxy matching procedure.
The $z_1$ descendent of the matched $z_2$
SA galaxy is identified as the counterpart of the
original SPH satellite. Again, this counterpart may be 
a satellite, a central galaxy, or (rarely) an `orphan'.
We find 40 out of 84 SPH satellite galaxies
at $z=0$ matched to SA satellites. Another 34 are matched to
SA central galaxies. The remaining 10 are unmatched, mostly because
they were formed under peculiar conditions in large halos
without being at the center of the halos. It is likely that 
many are associated with subhalos within the large halo, but our simulation
does not have adequate resolution to trace their formation
in detail. 

There are 8 SA satellites which are matched to SPH
central galaxies, and 74 SA satellites more massive than
$2\times 10^{10} h^{-1}M_{\odot}$ which are unmatched. A few of
the latter may, in fact, correspond to unmatched SPH satellites;
most correspond to systems in which the SPH gas did not cool.
The small number of SA satellites
matched to SPH centrals in comparison with the number of SPH
satellites matched to SA centrals is a consequence of the lower
effective resolution of the SPH simulation. We have checked that the
merging rate, estimated as the fraction of all satellites which 
merge between two output times, is actually very similar in the two
simulations.

\begin{figure}
\centering
\epsfxsize=\hsize\epsffile{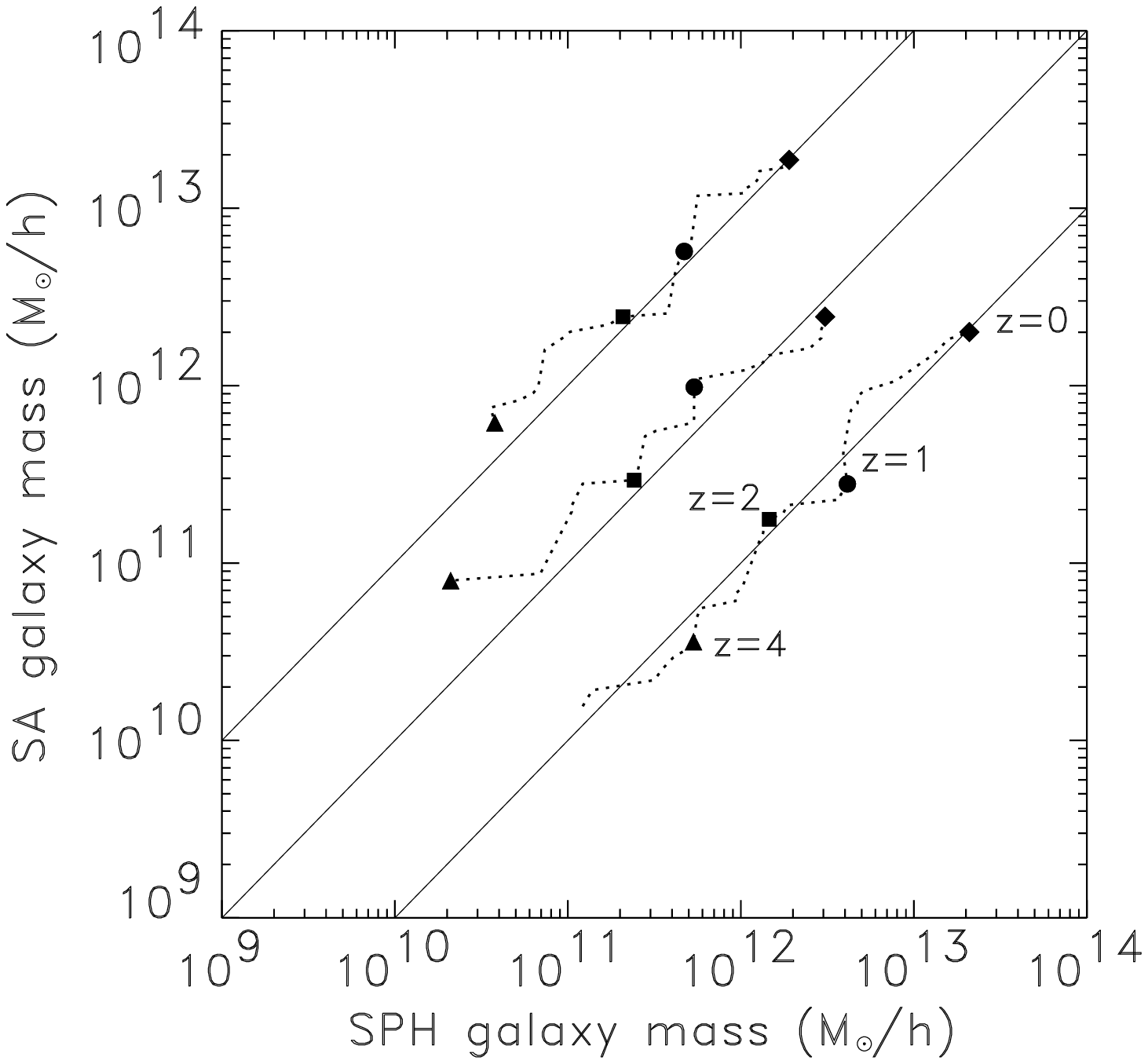}
\caption[The most massive galaxies' mass]
{We plot the mass evolution for the 3 most massive galaxies.
Dotted lines show their mass trajectories, and symbols
highlight the masses at 4 output redshifts; z=0 (diamond),
z=1 (circle), z=2 (square), and z=4 (triangle).  
Lines and symbols for the 3 galaxies are offset vertically 
by an order of magnitude each.}
\label{massevolution}
\end{figure}

Satellite-satellite matches are plotted as filled circles 
in Figure~\ref{fig_cool7}, whereas SA satellite -- SPH central and
SA central -- SPH satellite matches are plotted as open circles.
Unmatched satellites are plotted at an arbitrary small value of
the `missing' mass and so appear along the left and bottom edges
of the diagrams. The numbers of matches of each type are listed 
explicitly. The number given for the unmatched SA satellites
corresponds to objects with masses above $2\times 10^{10} 
h^{-1}M_{\odot}$, the minimum mass for an SPH galaxy.

For satellite--satellite matches the correspondence between the
masses found using our two techniques is still reasonably good
although with considerably more scatter than was the case for central
galaxies. There is a tendency for the SPH masses to be larger than
the SA masses suggesting that gas is continuing to cool onto the
galaxies after the time when they are last identified as central
objects. It should be noted that a group of points appearing in the
lower portion of the plot, indicating SPH satellites matched to 
small mass SA satellites, are mostly those which are matched by our particular
treatment for {\it forming} SPH galaxies, as explained above. 
In such cases, the halos hosting the forming galaxies merge  
into larger halos before the next output.
Their central SA galaxies then become satellites, and their masses 
stay constant, whereas the SPH galaxies grow in mass
continually until (and even after) they actually become satellites.
Unsurprisingly, for the satellite -- central galaxy matches 
the central galaxy mass is almost always substantially larger than 
the satellite galaxy mass, regardless of which is SPH and which SA.
It is noticeable that as one moves to higher redshift,
the number of unmatched SPH satellites increases substantially
relative to the number of unmatched SA satellites. This is probably
a reflection of the limited time resolution of our merger trees.
At early times structure is building up very rapidly and cooling
is efficient. The spacing between our outputs may then be too large
to resolve the formation and merging of significant numbers of halos.
These can form an SPH galaxy and turn it into a satellite between
two levels of our numerical merger tree. In such cases the SA scheme
will miss the satellite altogether. This highlights a different
resolution issue from those discussed above. The SA scheme must store
data sufficiently frequently that no significant stages of halo
formation are missed.

In general, we conclude from the overall level of agreement seen
for individual central and satellite galaxies that the
standard recipe of gas cooling in the semi-analytic model gives
an adequate description of the cooling process in the SPH model.

\section{Summary and discussion}
We have studied the cooling and condensation of gas onto protogalaxies
using two very different treatments of the gas physics within
the {\it same} simulation of dark matter evolution. One treatment uses
schematic recipes taken from semi-analytic galaxy formation models,
while the other tries to follow the hydrodynamics `correctly' and in
detail using an SPH approach. We have also tested
two different implementations of the SPH equations, 
confirming Springel \& Hernquist's (2001) conclusion that the outcome
of a simulation can depend sensitively on such details. In our case
the two SPH approaches lead to `galaxy' masses which differ by a
factor of 2, even for objects which are apparently well resolved.
We believe the energy and entropy conserving implementation of
Springel \& Hernquist (2001) to be the most reliable and have
concentrated on it when comparing SPH results to those of the
semi-analytic simulation.

We have characterised the effective resolution limit of simulations
using each of our gas modelling techniques by the lower limit
for which the simulated galaxy mass function agrees with that
of a much higher resolution simulation carried out using {\it the same}
technique. For the SA modelling this galaxy mass limit is just 
below the total baryon content of the least massive halo included in
construction of the merger trees (10 particles in the models used
here). For the SPH modelling the effective resolution limit is
significantly higher, particularly at low redshifts when cooling
times exceed the age of the Universe in most resolved halos.
For the simulations studied here the SPH mass resolution limit
at $z=0$ is roughly an order of magnitude larger than the SA limit, 
corresponding to a galaxy of about 75 SPH particles.

Above the resolution limit of both techniques (i.e. for galaxy
masses above $\sim 10^{11} h^{-1}M_{\odot}$) the $z=0$ galaxy
populations found by the two methods correspond remarkably well
on an object-by object basis. The agreement is particularly good
for the most massive galaxies in each halo (the `central' galaxies).
For satellite galaxies the agreement is still reasonably good
for objects which match between the two types of simulation, but
two significant types of discrepancy occur. One concerns objects which
merge with the central galaxy in one of the simulations but not in the
other. This is inevitable given the crudity of the recipe used to
determine merging probabilities in the SA model. The difficulty can be
avoided, at least for the more massive galaxies, with $N$-body simulations
of high enough resolution. The dark matter substructures associated 
with satellite galaxies can then be tracked explicitly and no SA recipe
is needed to follow their merging (see SWTK). 

A more interesting discrepancy is the significant number of SPH
satellites, which are unmatched in the SA simulation.
Some of these may reflect insufficient time resolution in the
simulation outputs from which the SA merger trees are constructed;
halos may collapse, form galaxies, and merge into a larger halo
between two of the stored outputs. Other unmatched SPH satellites may
reflect more fundamental failures of the SA scheme. A forming
dark halo may be sufficiently irregular to cool gas onto two
galaxies rather than one, or cooling shocks associated with
large-scale structure may fragment to make galaxies without any
associated halo. The number of unmatched SPH satellites is clearly
greater at early times than at $z=0$, suggesting that the process
producing these objects is more effective when cooling times are
short and structure growth is more rapid. This issue clearly warrants
further investigation. Although the unmatched SPH satellites are
a small population above the effective resolution limit of the SPH
simulation, they probably signal a real failure of the SA scheme.

It is instructive to compare the computational efficiency of our
two schemes. The comparison of this paper is based on implementing 
each of the schemes on top of a dark matter simulation with given
$N$. For the SA scheme the overhead in post-processing, although
non-negligible, is not dominant. Thus the computational cost is
essentially that of running the high resolution cosmological
$N$-body simulation. The SPH scheme requires an additional $N$
SPH particles, the force calculation for the SPH particles is
considerably more complex than for the dark matter particles, and
substantially shorter timesteps are needed to get convergent results
in an SPH simulation with cooling than in the corresponding pure
dark matter simulation. As a result, somewhat more 
than twice the memory and roughly an order of magnitude more
CPU time are needed to carry out the SPH simulation than the corresponding
pure $N$-body simulation. (The actual CPU factor for the simulations
of this paper was about 20, but in neither case were the
simulation parameters optimally chosen for the particular
configuration treated.) We have seen, however, that the effective
galaxy mass resolution of the SPH simulation is roughly an order of
magnitude worse than that of SA simulation. Thus an SPH simulation
with ten times as many particles would be needed to get
a galaxy population with the same effective mass resolution as
the SA simulation. This would require 20 times as much memory
and more than two orders of magnitude more CPU time than the
SA simulation.  

In practice these considerations imply that simulations can be
carried out using the SA technique which are far beyond the
current capabilities of the SPH technique. For example, the S4
cluster simulation of SWTK has 250 times better
mass resolution than the SA simulation of this paper and so of
order 2000 times better resolution than our SPH simulation. It
will be a long time before we can consider an SPH cluster/galaxy
formation simulation with 2000 times more particles than the 
one analysed above. A further advantage of the SA technique is that
the assumptions about uncertain aspects of the galaxy formation
modelling (e.g. the mode and efficiency of star-formation or feedback)
can be changed without any need to rerun the simulation. An example of
this was shown in Figure~\ref{massfunc}, where the predictions of the
high resolution S4 simulation in the absence of star-formation and
feedback could be calculated from the stored simulation output. 
Thus for simulations concentrating on the global properties of the
galaxy population and their dependence on the `subgrid' physics
(star formation, feedback, etc.) the SA technique is clearly the 
method of choice.

On the other hand, it is obvious that the SA technique can only be
as good as the recipes on which it is based. Our work has shown
that the standard gas cooling prescription works surprisingly well,
but has also uncovered some apparent discrepancies. Discrepancies
of this kind can only be studied and understood through comparison
with simulations (or analytic models) which attempt to follow the
relevant physics from first principles. Such models or simulations are
also needed, of course, to design the SA recipes in the first place.
In addition there are 
many aspects of galaxy formation where SA modelling can
only give crude indications. Confirmation through direct
simulation is then indispensable. Examples include the internal structure
of galaxies, the structure of the intergalactic medium, and the
comparison of both of these with the wealth of available data. 
Galaxy formation and its interaction with the environment
is a complex and highly nonlinear process involving many different
aspects of astrophysics. Gaining some understanding will require
concerted use of all the observational and theoretical weapons at
our disposal. SA and SPH simulation techniques will both remain
major parts of our theoretical arsenal.

\bigskip

We thank Simone Marri and Hugues Mathis for helpful comments.

\end{document}